\documentclass[fleqn,usenatbib]{mnras}
\usepackage[T1]{fontenc}
\usepackage{ae,aecompl}
\usepackage{graphicx}
\usepackage{amsmath}
\usepackage{amssymb}
\usepackage{url}
\graphicspath{{figure/}}

\newcommand{\kms}{\,km\,s$^{-1}$}
\newcommand{\molh}{$\text{H}_2$}
\newcommand{\Msun}{M$_\odot$}
\newcommand{\pc}[2]{pc$^{-2}$}

\newcommand{\Lya}{Ly$\alpha$}
\newcommand{\xion}{$\xi_\text{ion}$}
\newcommand{\Nion}{$\dot{N}_\text{ion}$}
\newcommand{\fesc}{$f_\text{esc}$}
\newcommand{\nion}{$\dot{n}_\text{ion}$}
\newcommand{\qhi}{$Q_\text{\ion{H}{I}}$}
\newcommand{\qhii}{$Q_\text{\ion{H}{II}}$}
\newcommand{\taucmb}{$\tau_\text{CMB}$}

\defcitealias{Yung2019}{Paper I}
\defcitealias{Yung2019a}{Paper II}
\defcitealias{Yung2020}{Paper III}

\defcitealias{Popping2014}{PST14}
\defcitealias{Somerville2015}{SPT15}
\defcitealias{Somerville2015a}{SD15}

\defcitealias{Bruzual2003}{BC03}
\defcitealias{Kennicutt1998}{KS}
\defcitealias{Gnedin2011}{GK}
\defcitealias{Bigiel2008}{Big}
\defcitealias{Kuhlen2012}{KF12}
\defcitealias{Becker2013}{BB13}
\defcitealias{Robertson2013}{R13}
\defcitealias{Robertson2015}{R15}

\title[SAM forecasts -- IV. Cosmic reionization]{Semi-analytic forecasts for \textit{JWST} -- IV. Implications for cosmic reionization and LyC escape fraction}

\author[L. Y. A. Yung et al.]{L. Y. Aaron\ Yung,$^{1,2}$\thanks{E-mail: yung@physics.rutgers.edu} 
Rachel S.\ Somerville,$^{1,2}$ Steven L.\ Finkelstein,$^{3}$ 
\newauthor Gerg\"{o}\ Popping,$^{4}$ Romeel\ Dav\'{e},$^{5,6,7}$ Aparna\ Venkatesan,$^{8}$ Peter\ Behroozi$^{9}$
\newauthor and Harry C.\ Ferguson$^{10}$
\\
$^{1}$ Department of Physics and Astronomy, Rutgers University, 136 Frelinghuysen Road, Piscataway, NJ 08854, USA\\
$^{2}$ Center for Computational Astrophysics, Flatiron Institute, 162 5th Ave, New York, NY 10010, USA\\
$^{3}$ Department of Astronomy, The University of Texas at Austin, Austin, TX 78712, USA\\
$^{4}$ European Southern Observatory, Karl-Schwarzschild-Strasse 2, D-85748 Garching, Germany\\
$^{5}$ Institute for Astronomy, University of Edinburgh, Edinburgh EH9 3HJ, UK\\
$^{6}$ Department of Physics and Astronomy, University of the Western Cape, Cape Town 7535, South Africa\\
$^{7}$ South African Astronomical Observatory, Cape Town 7925, South Africa\\
$^{8}$ Department of Physics and Astronomy, University of San Francisco, 2130 Fulton Street, San Francisco, CA 94117, USA\\
$^{9}$ Steward Observatory, University of Arizona, Tucson, AZ 85719, USA\\
$^{10}$ Space Telescope Science Institute, 3700 San Martin Drive, Baltimore, MD 21218, USA
}

\date{Accepted XXX. Received YYY; in original form ZZZ}

\pubyear{2020}
\begin{document}
\label{firstpage}
\pagerange{\pageref{firstpage}--\pageref{lastpage}}
\maketitle

\begin{abstract}
Galaxies forming in low-mass halos are thought to be primarily responsible for reionizing the Universe during the first billion years after the Big Bang. Yet, these halos are extremely inefficient at forming stars in the nearby Universe. In this work, we address this apparent tension, and ask whether a physically motivated model of galaxy formation that reproduces the observed abundance of faint galaxies in the nearby Universe is also consistent with available observational constraints on the reionization history. 
By interfacing the Santa Cruz semi-analytic model for galaxy formation with an analytic reionization model,
we constructed a computationally efficient pipeline that connects `ground-level' galaxy formation physics to `top-level' cosmological-scale observables.
Based on photometric properties of the galaxy populations predicted up to $z=15$, we compute the reionization history of intergalactic hydrogen. We quantify the three degenerate quantities that influence the total ionizing photon budget, including the abundance of galaxies, the intrinsic production rate of ionizing photons, and the LyC escape fraction. We explore covariances between these quantities using a Markov chain Monte Carlo method. We find that our locally calibrated model is consistent with all currently available constraints on the reionization history, under reasonable assumptions about the LyC escape fraction. 
We quantify the fraction of ionizing photons produced by galaxies of different luminosities and find that the galaxies expected to be detected in \textit{JWST} NIRCam wide and deep surveys are responsible for producing $\sim 40$--80\% of ionizing photons throughout the EoR.
All results presented in this work are available at \url{https://www.simonsfoundation.org/semi-analytic-forecasts-for-jwst/}.
\end{abstract}
\begin{keywords}
galaxies: evolution--galaxies: formation--galaxies: high-redshifts--galaxies: star formation--cosmology: theory--dark ages, reionization, first stars
\end{keywords}

\section{Introduction}
\label{sec:intro}

During the Epoch of Reionization (EoR), the intergalactic medium (IGM) underwent a global phase transition, during which the hydrogen progressively became ionized by the radiating Lyman-continuum (LyC) sources in the early universe \citep*{Miralda-Escude2000}.
Identifying and characterizing these sources remains a fundamental open challenge in modern cosmology. Indeed, this is one of the main science drivers of the \textit{James Webb Space Telescope} (\textit{JWST}). With the unprecedented infrared (IR) sensitivity and resolution of its on-board photometric instrument Near-Infrared Camera (NIRCam), \textit{JWST} is expected to detect many more faint galaxies during the EoR. In addition, \textit{JWST} will be able to provide additional constraints on the nature of the sources that reionized the Universe, such as revealing early accreting black holes. A number of planned \textit{JWST} observations, including both Guaranteed Time Observation (GTO), such as the \textit{JWST} Advanced Deep Extra-galactic Survey \citep[JADES;][]{Williams2018} and Early Release Science (ERS) projects, such as the Cosmic Evolution Early Release Science survey \citep[CEERS;][]{Finkelstein2017}, are designed to study and put constraints on the galaxy populations during the EoR, including both their statistical properties and the production rate of ionizing photons.

\subsection{The overall budget of ionizing photons}

It is clear that galaxies forming in the early universe have influenced large-scale events \citep{Dayal2018}. The cosmic ionizing photon budget is subject to three major moving parts, including the number density of galaxies, the intrinsic productivity of ionizing photons, and the LyC escape fraction.
The volume-averaged number density of high-redshift galaxies is only partially constrained by existing observations, which are limited by the sensitivity of existing facilities, particularly in the observed frame near-mid IR. 
To date, nearly 2000 galaxy candidates at $z \gtrsim 6$ have been detected in the Cosmic Assembly Near-IR Deep Extragalactic Legacy Survey \citep[CANDELS;][]{Grogin2011, Koekemoer2011}, \textit{Hubble} Ultra Deep Field \citep[HUDF;][]{Beckwith2006, Bouwens2011, Ellis2013, Oesch2013}, and UltraVISTA \citep{McCracken2012}, with faint objects of rest-frame UV luminosities reaching $M_\text{UV} \sim -17$ \citep[e.g.][]{Bouwens2015, Finkelstein2015}. 
Lensed surveys through massive foreground clusters can reach even fainter detection limits \citep[e.g.][]{Livermore2017, Lotz2017, Atek2018}, though this approach comes with high systematic uncertainties that remain poorly constrained \citep{Kawamata2016, Bouwens2017, Priewe2017}. As a result, there are still significant uncertainties on the faint-end slope of the UV luminosity functions (UV LFs) at $z \gtrsim 6$, which give rise to uncertainties of $\gtrsim 0.2$ dex on the integrated UV luminosity density at high redshift \citep{Ishigaki2018}, as well as the magnitude at which the UV LFs `turnover'.

The intrinsic production efficiency of ionizing radiation of high-redshift galaxies is subject to its own set of uncertainties. 
In early analytic calculations, this quantity was treated simply as a constant or as a parametrized function of redshift \citep{Madau1999, Finkelstein2012a, Kuhlen2012, Robertson2015, Mutch2016}. 
However, it is now recognized that this quantity depends strongly on many properties of the stellar populations in these early galaxies, including age, metallicity, upper mass cutoff of the stellar initial mass function (IMF), and binarity \citep{Eldridge2009, Eldridge2017, Topping2015, Wilkins2016a, Yung2020}.
There are still significant uncertainties in predictions of this quantity even in state-of-the-art stellar population synthesis (SPS) models \citep{Conroy2013}.
In general, we expect high-redshift galaxies to have younger, lower metallicity stellar populations, resulting in harder spectra yielding higher LyC production efficiencies. The contribution to the ionizing photon budget from sources such as X-ray binaries and Active Galactic Nuclei (AGN) also remains uncertain \citep[e.g.][]{Madau2017a, Manti2017}. Some recent studies have set out to constrain the production efficiency both locally and at high redshift using observations of UV-continuum slope, $\beta_\text{UV}$, H$\alpha$ and \ion{C}{IV} emission  \citep{Stark2015, Bouwens2016a, Schaerer2016, Shivaei2018, Emami2019, Lam2019}.

The fraction of ionizing radiation escaping to the IGM is the least constrained component among these three moving parts.
Simulations have shown that it is extremely sensitive to many detailed geometrical and physical features that act across many scales, including the internal distributions of dense gas, dust clouds, and stars within the interstellar medium (ISM) and the structure of the circumgalactic medium (CGM) \citep{Paardekooper2011, Paardekooper2013, Paardekooper2015, Benson2013, Kimm2014, Kimm2017, Kimm2019, Ma2015a, Xu2016, Popping2017a, Trebitsch2017, Trebitsch2018}.
These studies also found that the escape fraction does not correlate well with any particular global physical galaxy property and can scatter across an extremely wide range, from less than a thousandth to a few tens of a percent, even for galaxies of similar physical properties forming at the same epoch.
Many studies have attempted to constrain the escape fraction via observations and arrived at similar conclusions \citep[e.g.][]{Vanzella2010, Vanzella2015, Vanzella2018, Dijkstra2016, Guaita2016, Shapley2016, Grazian2017, Fletcher2019, Nakajima2020}.
Similar to the LyC production rate, many previous studies have treated the escape fraction as a single value \citep{Finkelstein2010, Finkelstein2012a, Finkelstein2015, Robertson2013, Robertson2015, Bouwens2015} or as a parametrized function of redshift or galaxy physical properties \citep{Wyithe2010, Kuhlen2012, Sharma2015, Naidu2020, Finkelstein2019}.

It is clear that detected galaxies alone are far from sufficient to reionize the universe \citep{Madau2008, Finkelstein2015, Robertson2015}. However, by assuming a LyC production efficiency and escape fraction that is consistent with that of bright galaxies, analytic calculations have shown that faint galaxy populations extrapolated from the observed UV LFs to below the current detection limits are able to provide the amount of ionizing photons needed to fully reionize the Universe in the required time-frame \citep{Finkelstein2012a, Finkelstein2015, Finkelstein2019, Kuhlen2012, Robertson2015, Bouwens2015a, Stark2016}.

\subsection{Constraints on the Epoch of Reionization}

The reionization history of the intergalactic hydrogen is constrained by a variety of IGM and CMB observations \citep{Fan2006}. 
During the phase transition, the depletion of neutral hydrogen along the line-of-sight can partially absorb high-redshift quasar spectra and leave behind a feature known as the Gunn-Peterson Trough \citep{Gunn1965, Becker2001, Fan2006a}.
The presence of intervening \ion{H}{I} also decreases the visibility of \Lya\ emitters, which puts a lower-limit to the redshift of the onset of the EoR  \citep{Stark2010, Dijkstra2011, Pentericci2011, Pentericci2014, Schenker2012, Schenker2014, Treu2013, Tilvi2014, Schmidt2016, Mason2018a}.
This same mechanism also enables the `Lyman-break selection' technique for identifying high-redshift galaxy candidates \citep{Steidel1996, Steidel1999}.
On the other hand, the CMB is scattered and polarized by free electrons in an ionized IGM. Therefore, the measured Thomson optical depth of the CMB, \taucmb, can be used to constrain the total number of electrons along the line of sight to the IGM.
The neutral IGM fraction towards the end of the EoR is constrained by a variety of observations (see \citealt{Robertson2015} for a concise summary).
Combining constraints on the onset and duration of the reionization process from various observations, the astronomical community has come to a general consensus that the phase transition of intergalactic hydrogen occurred approximately between $z=6$--10, and this period is often referred to as the Epoch of Reionization (EoR).

Historically, there has been tension among different observational constraints on the onset and duration of reionization. 
Early measurements of \taucmb\ reported by the Cosmic Background Explorer \citep[COBE;][]{Kamionkowski1994} and the Wilkinson Microwave Anisotropy Probe \citep[WMAP;][]{Spergel2003, Spergel2007, Komatsu2009, Komatsu2011, Hinshaw2013} seemed to imply a rapid reionization with a rather early conclusion \citep[e.g.][]{Somerville2003, Kuhlen2012, Robertson2013}.
On the other hand, a collection of \Lya\ forest constraints indicates that the number of ionizing photons reaching the IGM gradually flattens or even declines at $z\sim2$--6 (\citealt{Bolton2007, Faucher-Giguere2008a}, \citealt*{Prochaska2009}, \citealt{Songaila2010}). It was difficult to reconcile the early reionization apparently implied by the CMB (requiring a certain budget of ionizing photons) with the rather low emissivity at $z\sim 4$--6, while the galaxy population had presumably grown. One way to reconcile this tension was by invoking an `exotic' population of ionizing sources that contributed only at high redshift (such as Pop III stars or mini-quasars) or an escape fraction that strongly decreased with cosmic time. However, recent estimates of \taucmb\ reported by the Planck Collaboration (\citeyear{Planck2014, Planck2016, Planck2018}) have become considerably lower, indicating later reionization. At the same time, more recent work on the cosmic emissivity from \Lya\ forest constraints by \citet{Becker2013} indicates a higher emissivity towards the end of EoR at $z\sim 4$--6, largely alleviating the tension.  However, these measurements still provide important complementary constraints on the reionization history.

Another puzzle that has been discussed is the potential tension between the apparent need for relatively \emph{efficient} star formation in low mass halos at high redshift, needed to supply adequate numbers of the faint, low-mass galaxies that are invoked to make up the shortfall in the ionizing photon budget, and the much more \emph{inefficient} star formation in low-mass halos required to reconcile observed galaxy luminosity functions at low redshift with predicted halo mass functions in $\Lambda$ Cold Dark Matter \citep{Lu2014,Madau2014}. Observations of faint, low-mass galaxies in the nearby Universe provide important complementary constraints  to deep field studies on EoR populations \citep{Weisz2014, Boylan-Kolchin2014, Boylan-Kolchin2015, Boylan-Kolchin2016, Graus2016}.

\subsection{Current simulation efforts}

Modelling cosmic reionization is extremely challenging because, as we have outlined, it depends on accurately simulating structures from sub-pc scales to the largest structures in the Universe ($\sim 100$ Mpc). Several different complementary approaches have been presented in the literature. High-resolution cosmological zoom-in simulations, such as \textit{Renaissance} \citep{Wise2012, OShea2015}, FIRE \citep{Hopkins2014, Hopkins2018, Ma2018, Ma2017} and FirstLight \citep{Ceverino2019} simulate small volumes at relatively high resolution.
They are able to study the detailed properties of galaxies and their ISM, down to scales of tens of pc, but it is not feasible to simulate large volumes. Larger volume numerical hydrodynamic simulations such as EAGLE \citep{Schaye2015}, Illustris and IllustrisTNG \citep{Genel2014, Pillepich2018}, CROC \citep{Gnedin2014, Gnedin2014a, Gnedin2016, Gnedin2017}, CoDa \citep{Ocvirk2016, Ocvirk2018}, \textsc{BlueTides} \citep{Feng2016, Wilkins2017}, \textsc{Simba} \citep{Dave2019, Wu2020}, and \textsc{Sphinx} \citep{Rosdahl2018} are able to simulate larger volumes ($\sim$ 10--100 Mpc), but may not resolve the very low-mass halos that could be important for reionization, or the detailed properties of the ISM. A third approach is to simulate large volumes, in some cases with explicit modelling of radiative transfer in the IGM, but treating sources in a simplified way, e.g. by adopting empirical relations relating rest-UV luminosity to halo virial mass or stellar mass to estimate the number of ionizing photons produced \citep{Iliev2006, Iliev2006a, Trac2007, Trac2011, Santos2010, Hassan2016, Hassan2017}. This approach essentially operates under the same guiding principle that drives the popular (semi-)empirical modelling approach \citep{Behroozi2018, Moster2018, Tacchella2018, Finkelstein2019}.
However, this relies strongly on observational constraints, which must be extrapolated in regimes where these relations are not well calibrated.

The semi-analytic modelling approach is a middle way of bridging the gap between galaxy formation physics and the large-scale reionization history using physically motivated relationships between dark matter halo formation histories and galaxy properties. Semi-analytic models have had a long history of contributing to advancing the understanding of galaxy formation in ways that are complementary to numerical simulations \citep{Cole1994, Kauffmann1993a, Somerville1999, Somerville2015a}. They are grounded in the framework of dark matter halo `merger trees', and adopt simplified but physically motivated analytic recipes to model the main processes that shape galaxy formation. The models contain phenomenological parameters that are calibrated to reproduce a set of key observational relations in the nearby Universe. The models that we adopt here, the Santa Cruz Semi-analytic models \citep{Somerville1999,Somerville2008,Somerville2015}, have also been shown to reproduce a broad suite of other observations over a range of cosmic time and galaxy mass. The semi-analytic approach to studying reionization has also been adopted by the DRAGONS project \citep{Liu2016, Mutch2016, Geil2016}. Because of the computational efficiency of the semi-analytic approach, we are able to simulate large volumes down to the lowest mass halos that are expected to be able to cool via atomic cooling. In addition, we are able to explore variations in our model parameters. We have compared our model predictions with those from both high-resolution and large-volume cosmological hydrodynamic simulations \citep{Yung2019,Yung2019a}, and find excellent agreement.

In this series of \textit{Semi-analytic forecasts for JWST} papers, we have presented predictions for a variety of properties of high-redshift galaxy populations that are anticipated to be detected by \textit{JWST} or other future facilities. In \citet[hereafter \citetalias{Yung2019}]{Yung2019}, we presented distribution functions for the rest-frame UV luminosity and observed-frame IR magnitudes in \textit{JWST} NIRCam broadband filters. In \citet[hereafter \citetalias{Yung2019a}]{Yung2019a}, we further investigated the physical properties and the scaling relations for galaxies predicted by the same models. 
In \citet[hereafter \citetalias{Yung2020}]{Yung2020}, we made predictions for the intrinsic production rate of ionizing photons by high-redshift galaxies.
In this companion work (Paper IV), we combine our galaxy formation model with an analytic reionization model and a parametrized treatment of the escape fraction to explore the implications of our predictions for cosmic reionization. 
All results presented in the paper series will be made available at \url{https://www.simonsfoundation.org/semi-analytic-forecasts-for-jwst/}. We plan on making full object catalogues available after the publication of the full series of papers.

The key components of this work are summarized as follows:
the semi-analytic modelling pipeline, including the Santa Cruz galaxy formation model and the analytic reionization model are summarized briefly in Section \ref{sec:model}. 
Predicted reionization histories along with exploration of the effect of varying different model components are presented in Section \ref{sec:results}, including some specific predictions regarding \textit{JWST} in Section \ref{sec:jwst_obs}.
We discuss our findings in Section \ref{sec:discussion}, and a summary and conclusions follow in Section \ref{sec:snc}.

\section{The Modelling Framework}
\label{sec:model}

In this section, we present the components of a joint semi-analytic modelling pipeline for galaxy formation and cosmic reionization used to carry out this study. Throughout this work, we adopt cosmological parameters that are consistent with the ones reported by \citeauthor{Planck2016} in 2015: $\Omega_\text{m} = 0.308$, $\Omega_\Lambda = 0.692$, $H_0 = 67.8$\kms Mpc$^{-1}$, $\sigma_8 = 0.831$, and $n_s = 0.9665$. We adopt hydrogen and helium mass fractions $X = 0.75$ and $Y=0.25$.

\subsection{Semi-analytic model for galaxy formation}
\label{sec:sam}
The galaxy populations that source the ionizing photons are predicted using a slightly modified version of the well-established Santa Cruz SAM outlined in \citet*[hereafter \citetalias{Somerville2015}]{Somerville2015}.
We refer the reader to the following works for full details of the modelling framework: \citet{Somerville1999}; \citet*{Somerville2001}; \citet{Somerville2008}; \citet{Somerville2012}; \citet*[hereafter \citetalias{Popping2014}]{Popping2014} and \citetalias{Somerville2015}. For details on the model parameters used in this paper series and how they were calibrated, see \citetalias{Yung2019}.

The semi-analytic approach of modelling galaxy formation is based upon the merger histories of dark matter halos, sometimes referred to as `merger trees'. 
In this work, we adopted merger trees that are constructed using the Extended Press-Schechter (EPS) formalism \citep{Press1974, Lacey1993, Somerville1999a},
which have been shown to well-reproduce the statistical results for a large ensemble of merger trees extracted from $N$-body simulations \citep{Somerville1999a, Somerville2008, Zhang2008, Jiang2014}. 
This approach is able to achieve a wider dynamic range than any existing cosmological simulations, while requiring only a small fraction of computation resources.
For these reasons, our physical models are able to account for halos ranging from the very low-mass ones near the atomic cooling limit to the rare, massive ones across a wide range of redshift. The number density of `root' halos is computed based on results cosmological dark matter simulations \citep{Klypin2016, Rodriguez-Puebla2016,Visbal2018}. For further details, see \citet{Yung2019}.

Within these merger trees, SAMs then implement a set of coupled ordinary differential equations describing the flow of mass and metals between different components (diffuse intergalactic gas, hot halo gas, cold interstellar gas, the stellar body of the galaxy, etc). These flows are influenced by a range of physical processes, including cosmological accretion and cooling, star formation, chemical evolution, stellar-driven winds, and black hole feedback. The equations governing these processes contain `tunable' parameters that reflect our lack of a complete understanding of the basic physics. These parameters are calibrated to match a set of observational relationships at $z=0$. Note that in this paper series, as in all previous work with the Santa Cruz SAMs, \emph{the models have not been tuned to match observations at high redshift}.

The Santa Cruz model \citepalias{Popping2014, Somerville2015} includes a multiphase gas-partitioning recipe, which subdivides the cold gas content into an atomic, ionized, and molecular component, and a \molh-based stars formation recipe, which utilizes the predicted surface density of \molh\ ($\Sigma_\text{\molh}$) as a tracer for the surface density of SFR ($\Sigma_\text{SFR}$). In this work, we adopted the metallicity-based, UV-background-dependent partitioning recipe based on work by \citet[hereafter \citetalias{Gnedin2011}]{Gnedin2011} and the SF relation based on observations by \citet[hereafter \citetalias{Bigiel2008}]{Bigiel2008}. 
We note that recent evidence from both theory and observation suggests that the SF relation slope may steepen to $\sim 2$ at higher \molh\ surface densities \citep{Sharon2013, Rawle2014, Hodge2015, Tacconi2018}. 
In previous papers in this series, we have shown that this `two-slope' SF relation is crucial for our model to produce predicted galaxy populations that simultaneously match observational constraints on stellar mass, star formation rate, and rest-frame UV luminosity at $z=4$--10 \citep{Yung2019,Yung2019a}.
Thus, we refer to it as our fiducial model (\citetalias{Gnedin2011}-\citetalias{Bigiel2008}2).

\begin{figure*}
    \includegraphics[width=2\columnwidth]{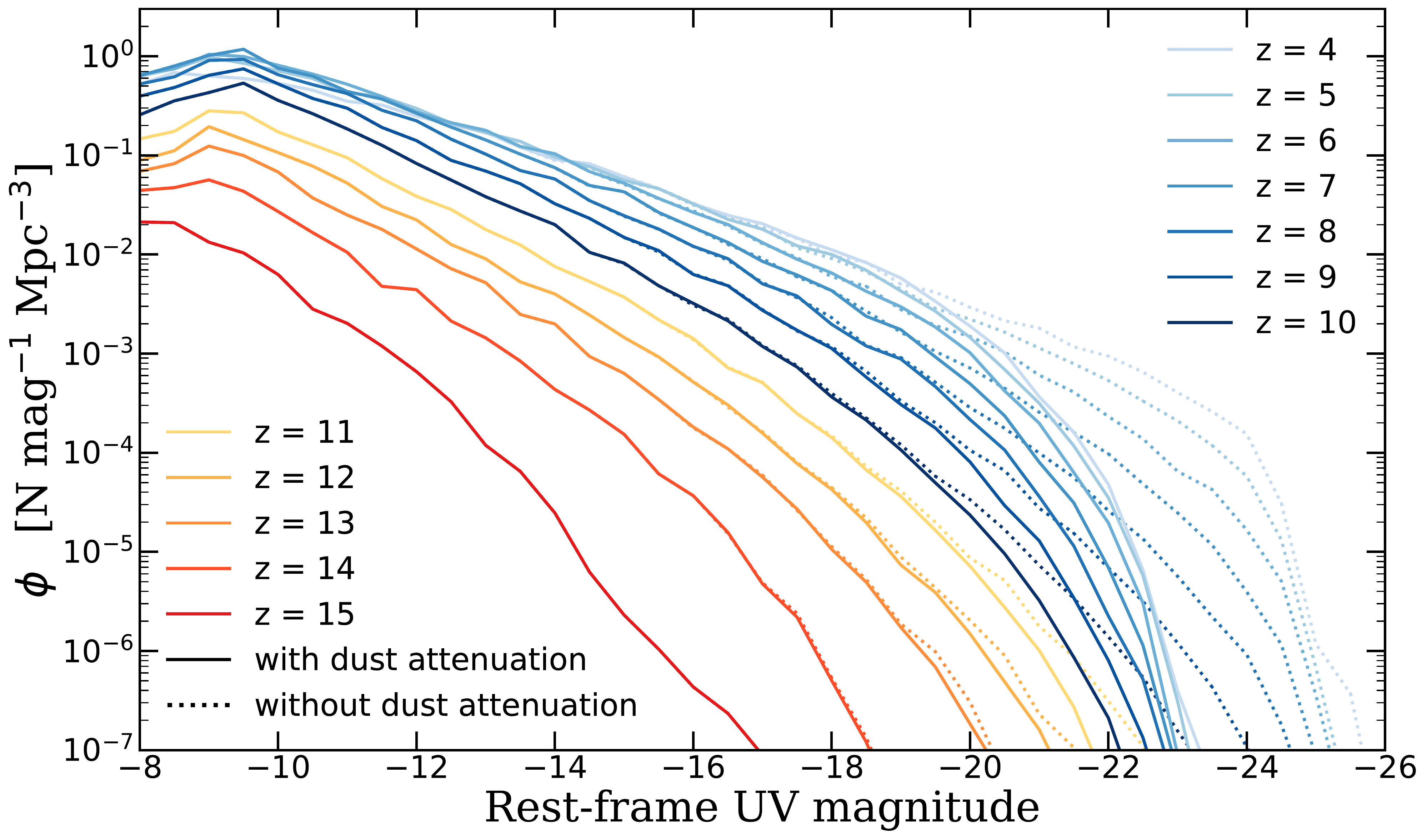}
    \caption{Predicted redshift evolution of the intrinsic (dotted) and dust-attenuated (solid) rest-UV LFs between $z = 11$ and 15 (this work; red colour series) and between $z=4$ and 10 (\citetalias{Yung2019}; blue colour series). The turnover at the faint end is not due to resolution, but rather to the atomic cooling limit. At very high redshifts, the UVLF is not well fit by a Schechter function. At $z\sim 4-8$, the apparent knee in the observed UVLF is largely due to differential dust extinction, which is larger in more luminous galaxies. }
    \label{fig:UVLFs_everything} 
\end{figure*}

The Santa Cruz SAM has been tested extensively in the past and shown to be able to reproduce a wide range of observables. In \citetalias{Yung2019}, the free parameters were re-calibrated to match a subset of $z \sim 0$ observations after adopting the updated cosmological parameters reported by the \citeauthor{Planck2016}. We then, in \citetalias{Yung2019} and \citetalias{Yung2019a}, identified the set of physical prescriptions (e.g. SF recipes) and physical parameters (e.g. SNe feedback slope) that are required to reproduce the evolution seen in observed galaxy populations up to $z\sim10$. 
This is encouraging as it suggests the physical processes that shape the formation of galaxies during reionization may not be so different from those that determine the properties of low-redshift galaxies.
Taking advantage of the model's efficiency, we also quantified the impacts on the predicted galaxy populations from the uncertainties in these model components by conducting controlled experiments where we systematically varied the model parameters. We found that the key process that has strong effects on the rest-frame UV luminosities and physical properties for bright, massive galaxies is the SF efficiency or timescale ($\tau_\text{*,0}$, see equation (1) in \citetalias{Yung2019}), which effectively characterizes the gas depletion time. For faint, low-mass galaxies, the UV LF is most sensitive to the stellar feedback relation slope ($\alpha_\text{rh}$, see equation (3) in \citetalias{Yung2019}), which characterizes the dependence of the mass loading factor of cold gas ejected by stellar feedback on halo circular velocity.
Currently there are not strong constraints on the faint-end slope of the UV LFs during EoR, where the predicted number density of faint galaxies across different models can vary by up to $\sim 1.5$ dex.

In the following subsections, we highlight how the main moving parts affecting the total emissivity of ionizing photons are treated in this work.

\subsubsection{Galaxy populations at ultrahigh redshifts}
\label{sec:211}

In order to quantify the contribution of ionizing photons from galaxies at ultrahigh redshifts ($z\gtrsim10$), we extend the predictions from our SAM up to $z\sim15$.
To assign a volume-averaged density to these galaxies, we use the same functional form for the HMF with the fitting parameters tuned to fit the results from the same set of simulations (Bolshoi-Planck and \citeauthor{Visbal2018}, see fig.~\ref{fig:HMF_ext}). See Appendix \ref{appendix:a} for full details and the values of all parameters. In fig.~\ref{fig:UVLFs_everything}, we present both the intrinsic (dust-free) and the dust-attenuated rest-frame UV luminosity functions predicted for the extended redshift range $z=11$--15.
In the same figure, we also compare these galaxies to the evolution between $z=4$--10 previously presented in \citetalias{Yung2019}. Tabulated values for the dust-attenuated UV LFs are provided in Appendix \ref{appendix:b}.
In Appendix \ref{appendix:c}, we compare the predicted UV LFs at $z=6$, 8, 9, and 10 to the latest observational constraints (obtained well after publication of our models) and find excellent agreement. We emphasize that the turnover at the faint end of our predicted luminosity functions is not due to resolution but is a result of the atomic cooling limit, which corresponds to a limiting halo mass that evolves with redshift.
We found in our predictions that the characteristic `knee' in UV LFs vanishes at $z \gtrsim 9$, seemingly due to both insignificant AGN feedback and lacking of dust (see fig.\ref{fig:UVLFs_everything}), and the faint-end slopes also gradually flatten as a function of rest-UV. The Schechter function is no longer a good representation and therefore, we do not provide Schechter fitting parameters.

\begin{figure}
    \includegraphics[width=\columnwidth]{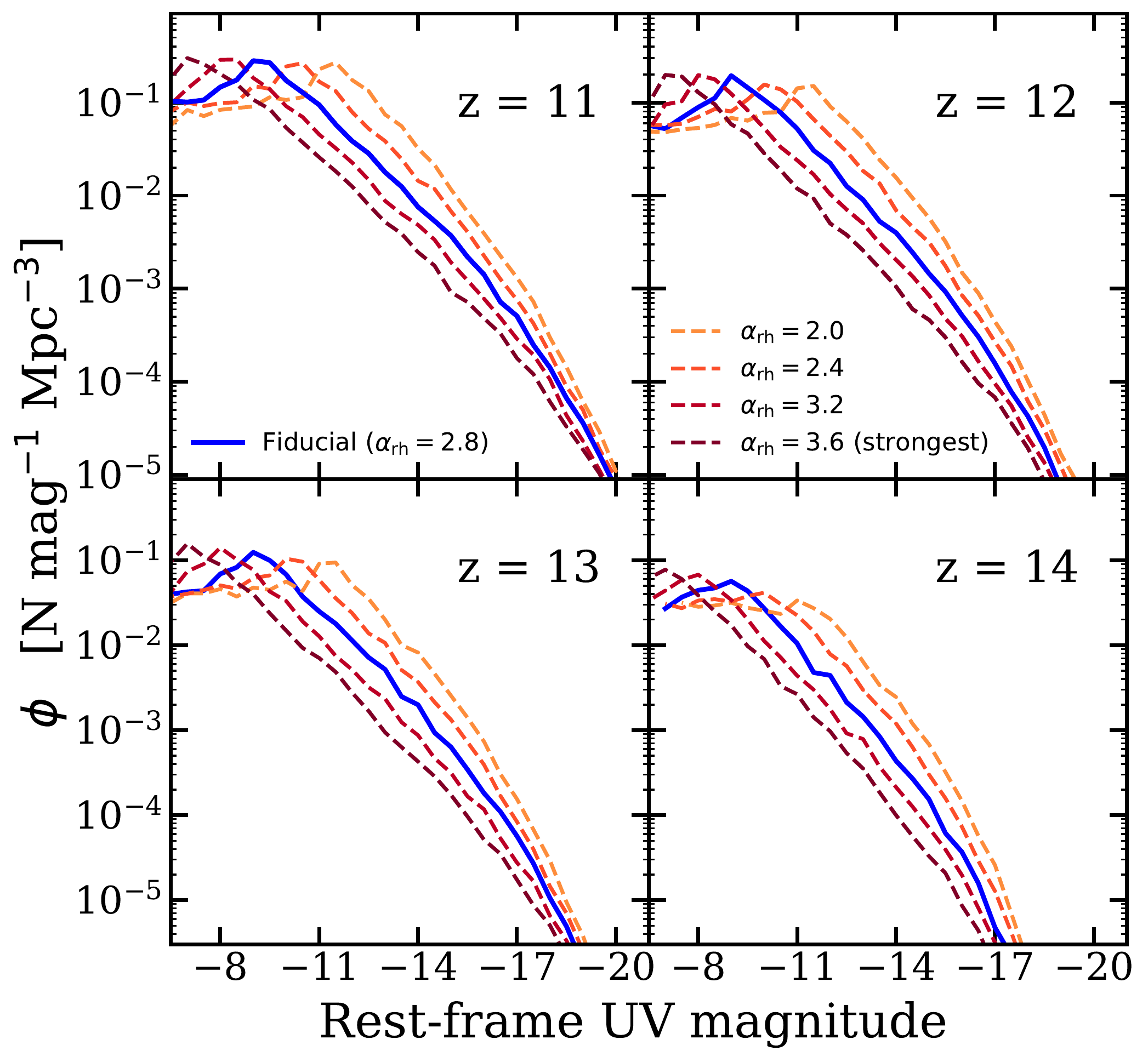}
    \caption{Redshift evolution of the dust-attenuated UV LFs between $z = 11$ and 14 predicted by our fiducial model ($\alpha_\text{rh} = 2.8$, blue solid line). We show four additional scenarios where we vary the parameter controlling the mass-loading of stellar driven winds, with $\alpha_\text{rh} = 2.0$ (weakest, lightest colour), 2.4, 3.2, and 3.6 (strongest, darkest colour). Larger values of $\alpha_\text{rh}$ produce stronger suppression of star formation in low-mass halos, leading to shallower faint end slopes and a lower luminosity for the turnover.}
    \label{fig:UVLFs_faint_ext} 
\end{figure}

We continue to explore the impacts from modelling uncertainties in the context of cosmic reionization. In fig.~\ref{fig:UVLFs_faint_ext}, we show UV LFs predictions for $\alpha_\text{rh} = 2.0$, 2.4, 3.2, and 3.6.
This is consistent with the findings in \citetalias{Yung2019} and \citetalias{Yung2019a}, where we showed that the faint-end slope of the UV LFs is inversely correlated with the stellar feedback parameter $\alpha_\text{rh}$ (i.e., a stronger dependence of wind mass loading on halo circular velocity leads to a flatter faint end slope). Furthermore, this effect also effectively shifts the halo occupation function and the turnover in the faint end of the UV LFs, which corresponds to the atomic cooling limit. In other words, we predict that the magnitude where the UF LF is truncated is inversely related with the strength of stellar feedback.

\begin{figure}
    \includegraphics[width=\columnwidth]{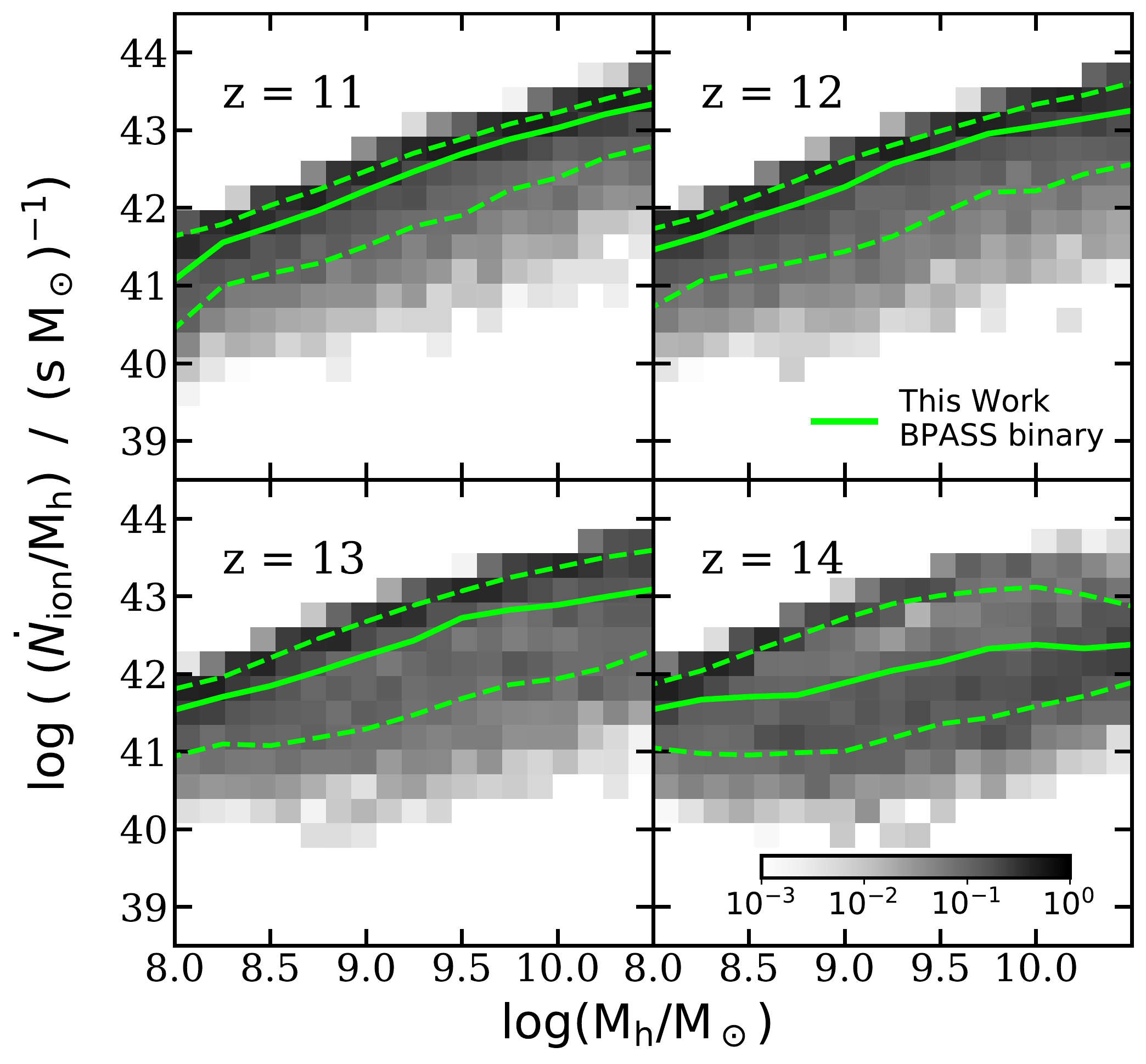}
    \caption{Specific ionizing photon production rate, $\dot{N}_\text{ion}/M_\text{h}$, as a function of halo mass between $z=11$ and 14, predicted by our fiducial model. The green solid and dashed lines mark the 50th, 16th, and 84th percentiles. The greyscale 2D histograms show the conditional number density per Mpc$^3$ in each bin, normalized to the number density in the corresponding (vertical) halo mass bin. The figure shows a decline in the specific ionizing photon production rate at fixed halo mass with increasing redshift, and a flattening dependence on halo mass. This is because these very early halos have not yet had time to form many stars. }
    \label{fig:NionMH_bin_ext} 
\end{figure}

\subsubsection{The intrinsic production rate of ionizing radiation}
\label{sec:212}

We refer to the ionizing photon production rate, \Nion, and the production efficiency, \xion, by stellar populations in galaxies, which does not account for the absorption or attenuation by the ISM and CGM, as `intrinsic'.
In \citetalias{Yung2020}, we self-consistently predict \Nion\ within the Santa Cruz modelling framework, based on the predicted star formation and chemical enrichment histories and results from stellar population synthesis (SPS) models.
This model component enables us to distinguish and track the contribution from galaxies across different rest-frame UV magnitudes and stellar masses.
In this work, we adopt the published results from the data tables released by the \citetalias{Bruzual2003}\footnote{\url{http://www.bruzual.org/~gbruzual/bc03/}} and the \textsc{bpass} group\footnote{\url{https://bpass.auckland.ac.nz/}, v2.2.1} \citep{Stanway2016, Eldridge2017, Stanway2018}.
Both models assume a Chabrier IMF with an upper mass cutoff $m_\text{U} = 100$\Msun.
These predictions for $z=4$--10 have been examined in detail in \citetalias{Yung2020}.
In that work, we also explored the scaling relations of \xion\ and \Nion\ with many SF-related physical properties and found that \xion\ is mildly correlated with $M_*$ and SFR, and these scaling relations evolve mildly as a function of redshift (where the underlying driving physical parameter is predominantly stellar metallicity).
Although the \textsc{bpass} SPS models account for mass transfer and mergers in stellar binaries, some processes that could potentially boost the production rate of ionizing photons, such as accreting white dwarfs and X-ray binaries, are not included in these models.

Here we extend these predictions to even higher redshift galaxy populations.
In Fig.~\ref{fig:NionMH_bin_ext}, we provide the predictions for the specific ionizing photon production rate, $\dot{N}_\text{ion}/M_\text{h}$, for halos in a relevant mass range at $z=11$--14.
The 2D histograms are shaded according to the conditional number density (Mpc$^{-1}$) of galaxies in each bin, which is normalized to the sum of the number density in its corresponding (vertical) halo mass bin.
The median, 16th, and 84th percentiles are marked in each panel to illustrate the statistical distribution.
Comparing to the predictions between $z=4$ to 10 shown in Fig. 7 in \citetalias{Yung2020}, which showed $\dot{N}_\text{ion}/M_\text{h}$ increases across the halo mass range explored, we find that the production rate per halo mass seem to have plateaued has noticeably larger scatter.

\subsubsection{Escape fraction of Lyman-continuum photons}
\label{sec:213}

The LyC escape fraction can be very stochastic depending on the many intricate physical processes occurring in individual galaxies and their internal structure. In this work, we take a simplistic approach and regard it as a \emph{population-averaged quantity}, which can either be understood as the population of galaxies all sharing the same escape fraction or as the escape fraction of the total number of ionizing photons collectively produced by all galaxies.  We treat it as a controlled free parameter, which may either be a constant value or evolve as a function of redshift.  For the remainder of this work, we refer to the LyC escape fraction as \fesc.

Inspired by the functional form presented by \citet{Kuhlen2012}, we adopt the following expression for the redshift-evolution of \fesc:
\begin{equation}
    f_\text{esc} (z) = \frac{f_\text{esc,max}}{1+\left(\frac{f_\text{esc,max}}{f_\text{esc,0}}-1\right)\;e^{-k_0(z-z_0)}} \text{,}
    \label{eqn:fesc}
\end{equation}
assuming \fesc\ decreases from some maximum value at high redshift, $f_\text{esc,max}$, at a characteristic growth rate, $k_0$, until it asymptotically reaches an anchoring valuing $f_\text{esc,0}$ at a given redshift $z_0 = 4$.
A goal of this work is to obtain constraints on \fesc\ under this empirical parametrization, as required by the set of currently available observational constraints.

\subsection{Analytic model for reionization history}
\label{sec:reion}

In this section, we present the set of analytic equations that tracks the reionization history of intergalactic hydrogen under the influence of the predicted galaxy populations. 
The model used in this work is similar to the ones presented in \citet[see also \citealt{Choudhury2009a, Finkelstein2012a, Finkelstein2019, Kuhlen2012, Shull2012,Robertson2015, Madau2017, Carucci2019, Naidu2020}]{Madau1999}, modified to fully utilize the predictions from the Santa Cruz SAM for galaxy formation. 
With this model, we can efficiently predict volume-averaged ionizing photon emissivity (\nion), IGM ionized fraction (\qhii), and the Thompson scattering optical depth (\taucmb).
In conjunction with the Santa Cruz SAM, the full modelling pipeline effectively connects the `ground-level' galaxy formation physics to the `top-level' cosmic reionization-related observables.
With this modelling pipeline, we explore and test the impact of individual model components and how they impact the cosmological scale observables. 
Note that predictions for helium reionization are beyond the scope of this work.

\subsubsection{Ionized volume fraction}

The temporal evolution of the volume-averaged ionizing volume filling fraction of ionized hydrogen, $Q_\text{\ion{H}{II}}$, is described by the first-order differential equation 
\begin{equation}
    \frac{dQ_\text{\ion{H}{II}}}{dt} = \frac{\dot{n}_\text{ion}}{\bar{n}_\text{H}} - \frac{Q_\text{\ion{H}{II}}}{\bar{t}_\text{rec}} \text{,}
    \label{eqn:reionization}
\end{equation}
derived in \citet{Madau1999}. The two terms can be interpreted as a growth term and a sink term, respectively, where the former is the ratio of the comoving ionizing emissivity, $\dot{n}_\text{ion}$, and the volume-averaged comoving number density for intergalactic hydrogen, $\bar{n}_\text{H}$; the latter is characterized by the ionized volume fraction divided by the recombination timescale of ionized hydrogen, $\bar{t}_\text{rec}$. We adopted $\bar{n}_\text{H} = 1.9 \times 10^{-7}$ cm$^{-3}$ as reported by \citet{Madau2014}.

\begin{table*}
    \centering
    \caption{Summary of components for reference model.}
    \label{table:modelcomponents}
    \begin{tabular}{llll}
        \hline
        Model / Constraints & References & Configurations & Remarks \\
        \hline
        Star formation & \citealt{Bigiel2008} & two-slope ($1\rightarrow2$)& adopted as implemented in \citetalias{Popping2014} and \citetalias{Somerville2015} \\
        Gas partitioning & \citealt{Gnedin2011} & metallicity-based & adopted as implemented in \citetalias{Popping2014} and \citetalias{Somerville2015} \\
        Stellar feedback & \citealt{Somerville2008}& $\alpha_\text{rh} = 2.4$, $\varepsilon_\text{SN} = 1.7$ & re-calibrated and tested in \citetalias{Yung2019} and \citetalias{Yung2019a} \\
        LyC productivity & \citealt{Stanway2018} & binary, v2.2.1 & newly implemented and tested in \citetalias{Yung2020} \\
        \ion{H}{II} recombination & \citealt{Hui1997} & Case B & adopted as implemented in \citealt{Finkelstein2019} \\
        \ion{H}{II} clumping factor & \citealt{Pawlik2015} & L25N512 simulation & adopted as implemented in \citealt{Finkelstein2019} \\
        \hline
        Emissivity constraints & \citealt{Becker2013} & \nion\ at $z=2$--5 & derived with cosmology consistent with this work \\
        CMB constraints & \citealt{Planck2016a}& $\tau_\text{CMB} = 0.058\pm0.012$ & derived with cosmology consistent with this work \\
        \hline
    \end{tabular}
\end{table*}

\subsubsection{Ionizing emissivity}

The comoving emissivity of ionizing photons, \nion, is the total budget supplied to reionize the IGM by galaxies, which is commonly modelled as the product of cosmic SFR or UV density, the LyC production efficiency of ionizing photons, and the fraction of photons that escapes to the IGM
\begin{equation}
    \dot{n}_\text{ion} = f_\text{esc}\;\xi_\text{ion}\;\rho_\text{UV} \textit{.}
\end{equation}
Recalling that in our models, $f_\text{esc}$ and $\xi_\text{ion}$ may have a different value for each galaxy, instead of combining these for the whole population as above, we calculate the comoving value at each redshift by summing over all predicted galaxies
\begin{equation}
    \dot{n}_\text{ion} = \sum\limits_{i}\; n_{\text{h}, i}\;\dot{N}_{\text{ion},i}\;f_{\text{esc},i} \text{,}
    \label{eqn:nion}
\end{equation}
where $n_\text{h}$ is the number density per Mpc$^3$ for each galaxy $i$, assigned based on the virial mass of the host halo (section \ref{sec:211}), \Nion\ is the intrinsic ionizing photon production rate (section \ref{sec:212}), and \fesc\ is the LyC escape fraction (section \ref{sec:213}).
This modified approach  does not require a predetermined truncation value of $M_\text{UV}$, as the turnover in the galaxy UV LF is a physical feature of our model. 
Moreover, Fig.~7 of \citetalias{Yung2019} and Fig.~\ref{fig:UVLFs_faint_ext} have shown that the magnitude where the UV LF turns over is directly correlated with the faint-end slope, which are both affected by the SN feedback slope $\alpha_\text{rh}$.
Therefore, instead of exploring a range of LF faint-end slope as is frequently done in other studies, we explore a range of $\alpha_\text{rh}$.

\subsubsection{Intergalactic \ion{H}{II} recombination timescale}

The recombination timescale for intergalactic hydrogen is given by
\begin{equation}
    \bar{t}_\text{rec} = \left[C_\text{\ion{H}{II}}\; \alpha_\text{B}(T)\; (1+\eta Y/4X)\; \bar{n}_\text{H}\;(1+z)^3 \right]^{-1} \text{,}
    \label{eqn:t_rec}
\end{equation}
where $C_\text{\ion{H}{II}}$ is a redshift-dependent \ion{H}{II} clumping factor and $\eta = 1$ for singly ionized helium at $z > 4$. We adopted numerical predictions for the clumping factor from the radiation-hydrodynamical simulation L25N512 by \citet*{Pawlik2015}, which $C_\text{\ion{H}{II}}$ evolves from $\sim1.5$ to $\sim4.8$ between $z\sim14$ to $\sim6$.
The quantity $\alpha_B(T)$ is the temperature-dependent case B recombination coefficient for hydrogen given in \citet{Hui1997}, where we adopt $T = 2\times 10^4$ K for the temperature of the IGM at the mean density; $\bar{n}_\text{H}$ is the mean density of hydrogen in the IGM.
Under the limitation of this type of model, we assume homogeneous recombination.

\subsubsection{Thompson scattering optical depth of the CMB}

The reionization history, $Q_\text{\ion{H}{II}}(z)$, is obtained by solving eqn.~\ref{eqn:reionization} using Python tools from \texttt{scipy.integrate.odeint} and \texttt{astropy.cosmology} \citep{Robitaille2013, Price-Whelan2018, Virtanen2020}.
We can then calculate the Thomson scattering optical depth of the CMB, $\tau_\text{CMB}$, using
\begin{equation}
    \tau_\text{CMB} = \int^{\infty}_{0} dz\frac{c(1+z)^2}{H(z)}Q_\text{\ion{H}{II}}(z) \sigma_\text{T} \bar{n}_\text{H} (1 + \eta Y/4X) \text{,}
    \label{eqn:tau_cmb}
\end{equation}
where $H(z)$ is the Hubble constant and $\sigma_\text{T}$ is the Thomson cross section.

The main components of our default `reference model', used throughout the remainder of this work are summarized in Table \ref{table:modelcomponents}.
This approach provides quick estimates of the volume-averaged reionization history and other cosmological-scale observables. However, it does not track the growth of individual Stromgren spheres. 
It also does not account for local density variances (e.g. void or over-dense regions), which may significantly affect the reionization histories on small scales.
We will further discuss the limitations of the model in section \ref{sec:discussion}.

\section{IGM reionization by high-redshift galaxies}
\label{sec:results}

In this section, we present a collection of predicted reionization histories and investigate how galaxy formation physics can affect these predictions. 
We experiment with a range of constant values of \fesc\ (section \ref{sec:constant}) or treat it as a function of redshift (section \ref{sec:variable}).
We also present a comparison with two other analogous studies (section \ref{sec:model_compare}).
At the end of the section, we probe the contribution of galaxies from different rest-$M_\text{UV}$, as well as forecasting the contribution from galaxies observable by \textit{JWST}.

\begin{figure*}
    \includegraphics[width=2\columnwidth]{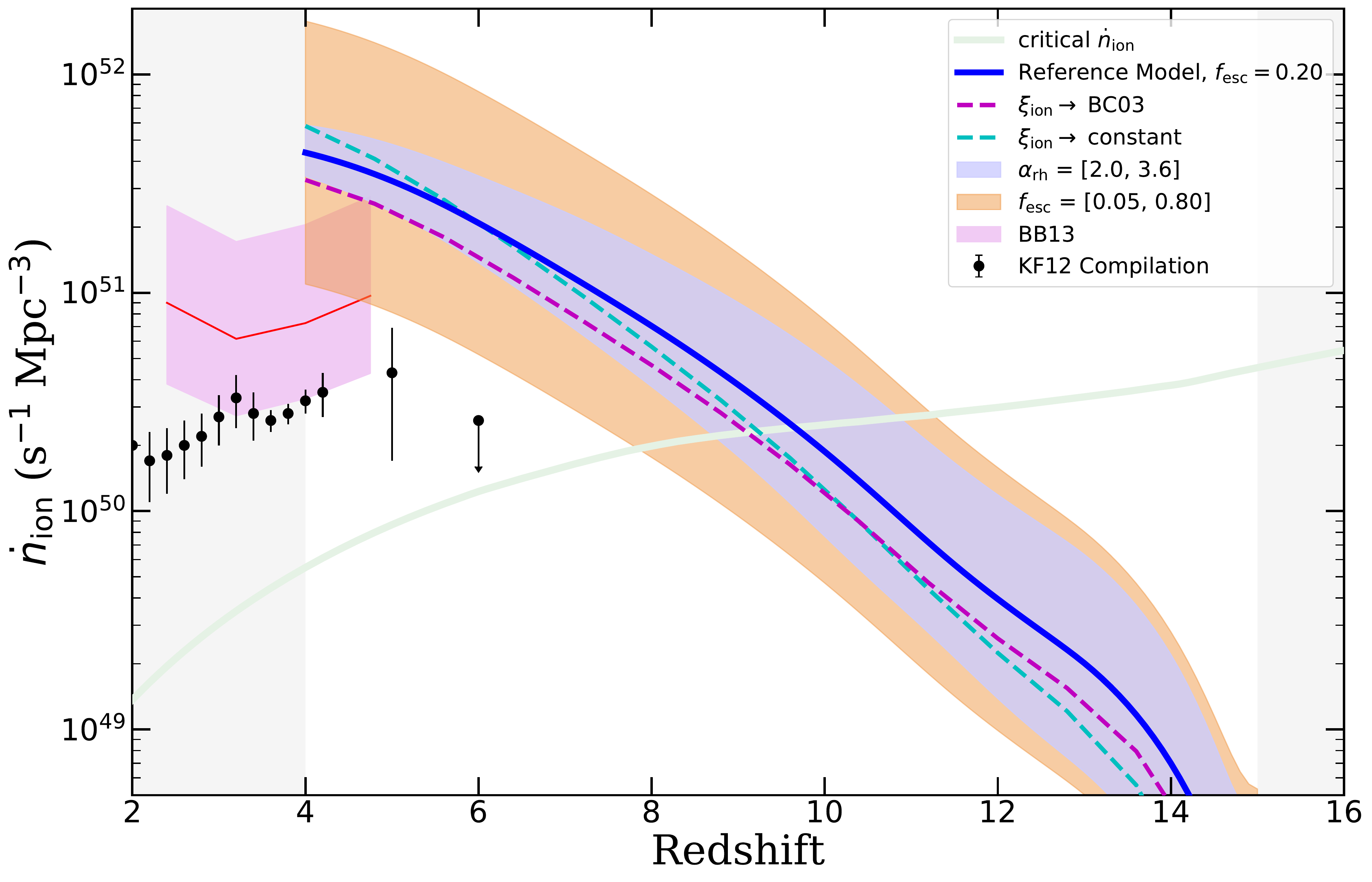}
    \caption{Ionizing photon emissivity, \nion, as a function of redshift predicted with the reference model configurations (blue solid, see Table \ref{table:modelcomponents}). The blue shaded region marks the range of \nion\ from galaxy populations predicted for $\alpha_\text{rh} = 2.0$ (weaker feedback, leading to a higher number density of low-mass galaxies and higher emissivity) and  3.6 (stronger feedback, leading to fewer low-mass galaxies and lower emissivity). The orange region marks the range predicted with $f_\text{esc}=0.05$ (lower emissivity due to low escape fraction) and 0.80 (higher emissivity). We also include predictions made with a constant $\log(\xi_\text{ion})=25.30$ and with \Nion\ from the SPS models of \citetalias{Bruzual2003}. These results are compared to observational constraints from \citetalias{Becker2013} and a compilation from \citetalias{Kuhlen2012}. The light green band shows the critical ionizing photon emissivity required to keep the Universe ionized (see eqn. \ref{eqn:t_rec} and associated description in text). This shows how uncertainties in different model components could have affected the total ionizing budget throughout the EoR.}
    \label{fig:unified_nion}
\end{figure*}

\begin{figure*}
    \includegraphics[width=1.8\columnwidth]{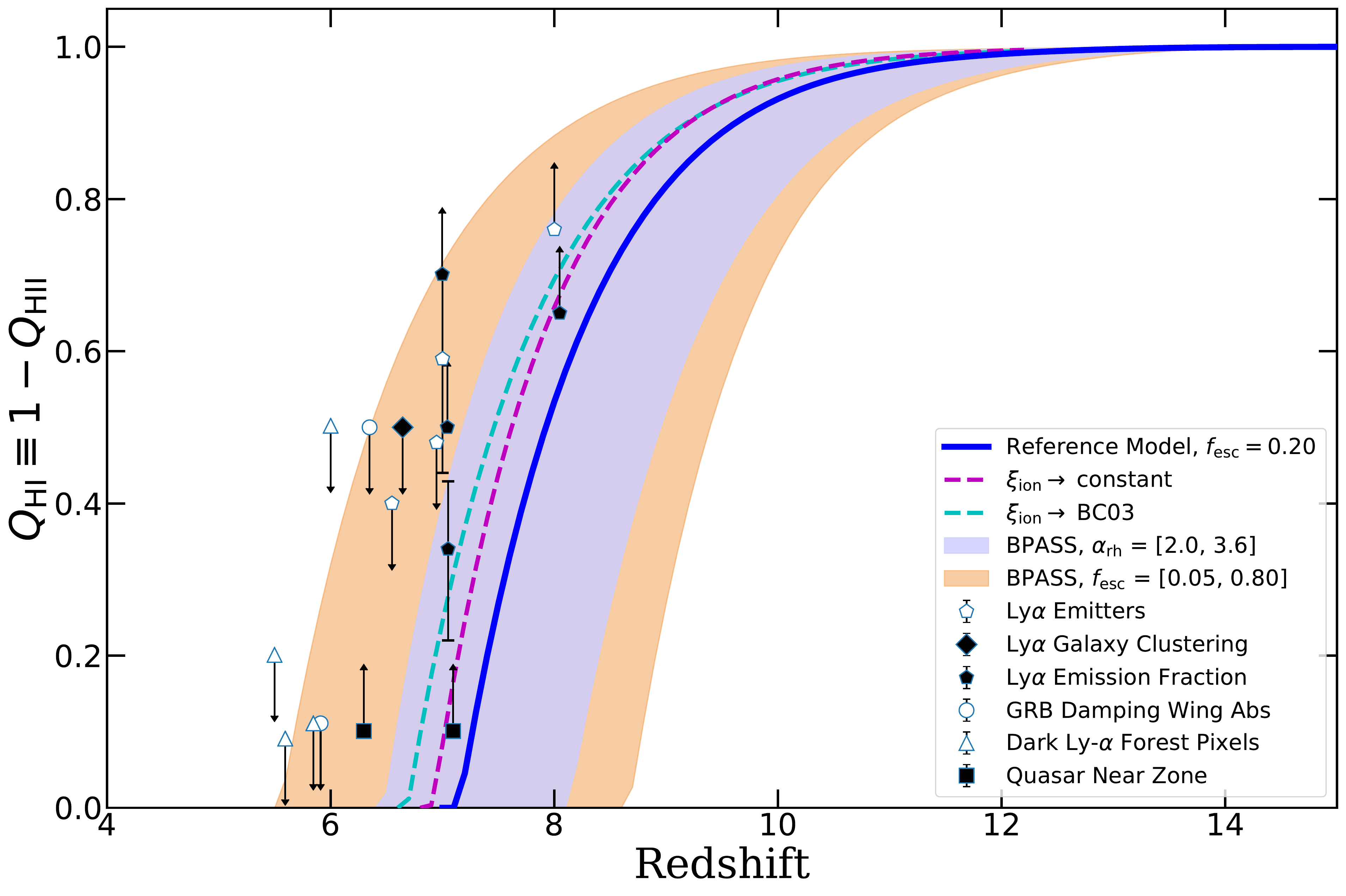}
    \caption{Neutral volume fraction, $Q_\text{\ion{H}{I}} \equiv 1 - Q_\text{\ion{H}{II}}$, as a function of redshift calculated for our reference model (blue solid). The blue shaded region marks the range for galaxy populations predicted for $\alpha_\text{rh} = 2.0$ (weaker feedback; lower bound) and  3.6 (stronger feedback; upper bound). The orange region marks the range predicted with $f_\text{esc}=0.05$ (upper bound) and 0.80 (lower bound). We also include predictions made with a constant $\log(\xi_\text{ion})=25.30$ and with \Nion\ from \citetalias{Bruzual2003}. These results are compared to a compilation of observational constraints from \citetalias{Robertson2013} and \citetalias{Robertson2015}, and additional constraints from \citet{Mason2018, Mason2019}. The simple reference model, with fixed \fesc, is in tension with these observations, though it is clear that this is primarily due to uncertainties on \fesc. }
    \label{fig:unified_qhii}
\end{figure*}

\subsection{Constant escape fraction}
\label{sec:constant}
At first, we take the simplest approach by letting \fesc\ be a non-evolving, universal quantity.
We present predictions for $\dot{n}_\text{ion}$, $Q_\text{\ion{H}{II}}$, and $\tau_\text{CMB}$ using our reference model configurations.
Taking advantage of the efficiency of our modelling pipeline, we then performed a controlled experiment by varying a set of selected model components to quantity their impact on these predictions. 

In Fig.~\ref{fig:unified_nion}, we show the evolution of \nion\ predicted by the reference model assuming $f_\text{esc} = 0.20$.
These results are compared to constraints on the global LyC emissivity at $2 < z < 5$ derived from the high-redshift \Lya\ forest by \citetalias{Becker2013}. 
The plotted data are for the fiducial temperature-density parameter $\gamma=1.4$ and spectral index of the ionizing sources $\alpha=2.0$, with shaded area showing the reported total error. 
Historically, there has been tension between the Thomson scattering optical depth of the CMB, $\tau_\text{CMB}$ and the ionizing photon emissivity at intermediate redshift, as discussed in the introduction. 
To demonstrate how these new constraints have eased the tension, we also show the compilation of constraints presented in \citet{Kuhlen2012}, which includes \Lya\ forest observations from \citet{Bolton2007, Faucher-Giguere2008, Prochaska2009, Songaila2010}.
The \citetalias{Becker2013} constraints are a factor of $\sim2$ higher than the previous measurements, and no longer require the total LyC emissivity to decrease so rapidly at $z\lesssim6$.
We also show the critical comoving ionizing emissivity, $\dot{n}^\text{crit}_\text{ion}$, or the minimum $\dot{n}_\text{ion}$ that is required to keep the Universe ionized
\begin{equation}
    \dot{n}^\text{crit}_\text{ion} = C_\text{\ion{H}{II}}\; \alpha_\text{A}(T)\; (1+\eta Y/4X)\; {\bar{n}_\text{H}}^2\;(1+z)^3 \textit{,}
\end{equation}
obtained by inverting the recombination timescale given in eqn.~\ref{eqn:t_rec}. 
Here, the temperature-dependent case A recombination coefficient for hydrogen, $\alpha_A(T)$, given by \citet{Hui1997} is invoked because direct recombination from free to the ground bound state is more likely to occur in the optically thin IGM. On the contrary, case B is more suitable at describing regions near a source given that photons released by free-to-ground recombination are likely to reionize a nearby hydrogen atom in these denser regions (see \citealt{Faucher-Giguere2009} for an in-depth discussion).
The rest of the variables are consistent with the ones adopted in our calculation of the \ion{H}{II} recombination timescale.

We compare these results with alternative scenarios predicted with a range of $f_\text{esc}$ and SN feedback slopes $\alpha_\text{rh}$.
As previously explored in the series, we found that SN feedback is the dominant process that regulates star formation in low-mass halos.
Deviating from the fiducial value $\alpha_\text{rh} = 2.8$, we found that the range $\alpha_\text{rh} = 2.0$ to 3.6 yield a range of faint-end slopes that are still well within the current observational uncertainties at $z \gtrsim 6$.
As shown in Fig.~7 of \citetalias{Yung2019} and in Fig.~\ref{fig:UVLFs_faint_ext}, low-mass galaxies are more abundant when feedback is weaker ($\alpha_\text{rh} = 2.0$) and, conversely, less abundant when feedback is stronger ($\alpha_\text{rh} = 3.6$).
Here, we show the range of predicted \nion\ for galaxy populations predicted with these boundary cases and found that these yield results nearly $\sim1$ dex apart.
Similarly, we also experiment with a wide range of $f_\text{esc} = 0.05$ to 0.80 to quantify its impact on the overall emissivity. 
From these results, we can already see that the LyC emissivity is more sensitive to the escape fraction than the faint-end slope of the UV LFs, for these variables within a physically meaningful range. 

To explore the uncertainties in modelling \Nion, we added predictions with \Nion\ from \citetalias{Bruzual2003}, which is the least optimistic model explored in \citetalias{Yung2020}, and a scenario with constant \xion\ adopting the expression from \citet{Kuhlen2012}
\begin{equation}
    \dot{N}_\text{ion} = 2\times10^{25} \text{s}^{-1} \left(\frac{L_\text{UV}}{\text{erg s}^{-1} \text{Hz}^{-1}}\right) \zeta_\text{ion} \text{,}
\end{equation}
where $\zeta_\text{ion}$ is a free parameter that characterizes the hardness of the spectra.
Here we assumed $\zeta_\text{ion} = 1$ as in the fiducial model of \citetalias{Kuhlen2012}. The rest-UV magnitude is converted using $\log_{10}(L_\text{UV}/(\text{erg s}^{-1} \text{Hz}^{-1})) = 0.4(51.63-M_\text{UV})$. 
This is equivalent to adopting a constant $\log{\xi_\text{ion}} = 25.30$. 
We find that models with \Nion\ computed self-consistently from the SPS models result in a shallower growth in \nion\ over time comparing to the model with a constant \xion. This is likely due to ageing and metal enrichment in the stellar populations in these galaxies, which naturally make the production of ionizing photons less efficient, although the number density of galaxies is growing.
However, this effect is insufficient to reproduce the \citetalias{Becker2013} constraints as the flattening due to changes in $\xi_\text{ion}$ is quite subtle, and overall \nion\ is still largely dominated by the fairly rapid growth of the overall number density of galaxies.
On the other hand, results using \Nion\ predicted by different SPS models seems to have evolved quite similarly over time with the expected factor of $\sim 2$ offset due to the inclusion of binary stars in the \textsc{bpass} models.
For further discussion of differences between the \textsc{bpass} binary SPS model and \citetalias{Bruzual2003}, we refer the reader to the discussion associated with Fig. 12 in \citetalias{Yung2020}.

\begin{figure}
    \includegraphics[width=\columnwidth]{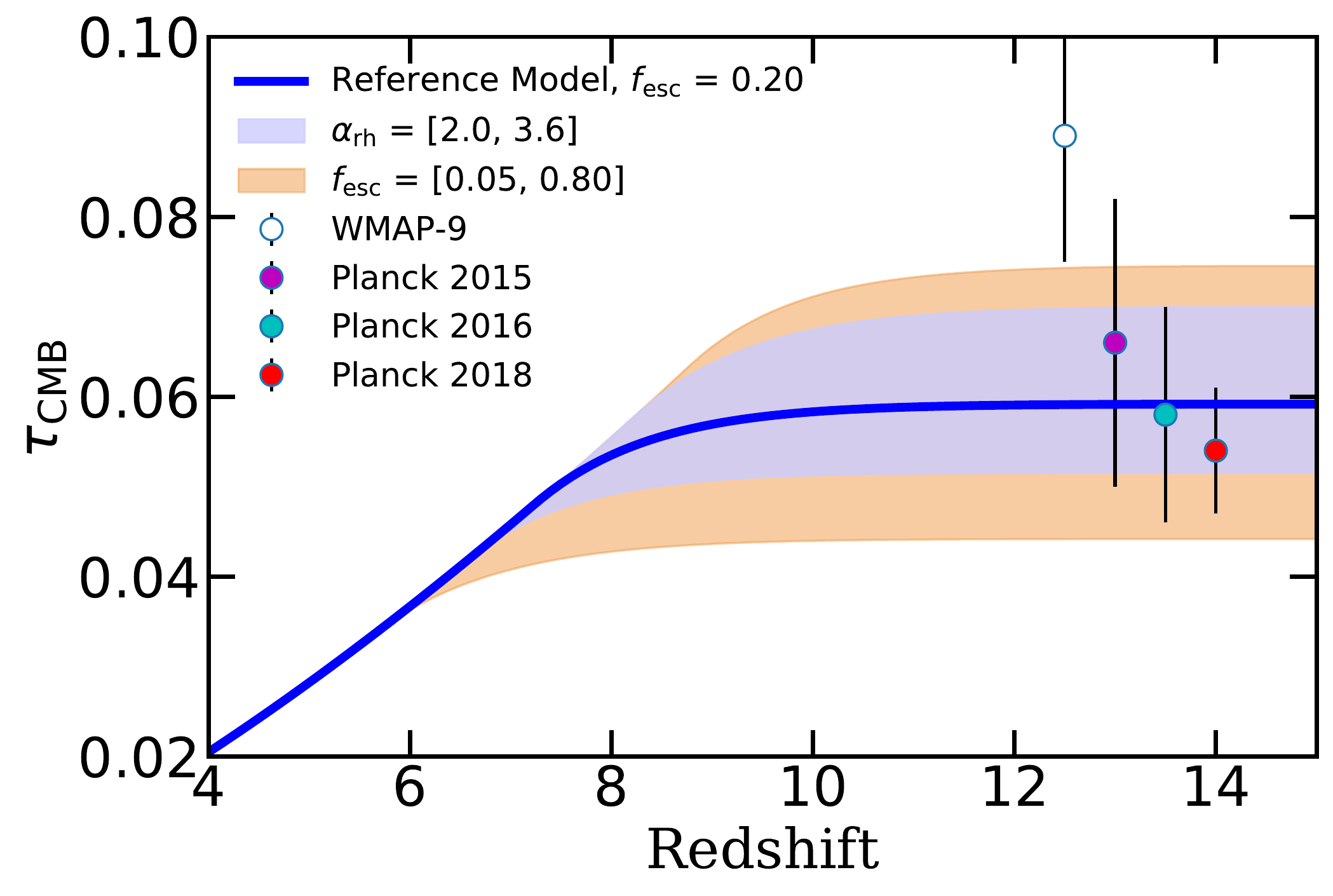}
    \caption{Thomson scattering optical depth of the CMB, \taucmb, as a function of redshift predicted by our reference model (blue solid). The blue shaded region marks the range of integrated \taucmb\ from galaxy populations predicted for $\alpha_\text{rh} = 2.0$ (upper bound) and  3.6 (lower bound). The orange region marks the range predicted with $f_\text{esc}=0.05$ (lower bound) and 0.80 (upper bound). We also show reported values from the Planck Collaboration \citeyear{Planck2014, Planck2016a, Planck2018} and from WMAP-9 \citep{Hinshaw2013}. Our reference model has no difficulty accounting for the more recent estimates of \taucmb.}
    \label{fig:unified_tau_z}
\end{figure}

In the same spirit as the \nion\ comparison, in fig.~\ref{fig:unified_qhii} we present the predicted IGM neutral fraction, $Q_\text{\ion{H}{I}} \equiv 1- Q_\text{\ion{H}{II}}$, from the same set of model variants. 
These predictions are stacked up against a compilation of observational constraints compiled from \citetalias{Robertson2013} and \citetalias{Robertson2015}, which consist of various kinds of observations, including \Lya\ emitting galaxies \citep{Ota2008, Ouchi2010, Pentericci2014, Schenker2014}, \Lya\ emission fraction \citep{McQuinn2007, Mesinger2008, Dijkstra2011}, \Lya\ galaxy clustering \citep{Ouchi2010}, \Lya\ damping wing \citep{Totani2006, McQuinn2008, Chornock2013}, from the near zones of bright quasars \citep{Bolton2007, Bolton2011, Schroeder2013}, and from dark pixels in \Lya\ forest measurements \citep{Mesinger2010, McGreer2011, McGreer2015}.  We refer the reader to \citet{Robertson2013} for a detailed description of these constraints. We also added the latest constraints from \Lya\ emitting galaxies reported by \citet{Mason2018, Mason2019}. 

\begin{figure}
    \includegraphics[width=\columnwidth]{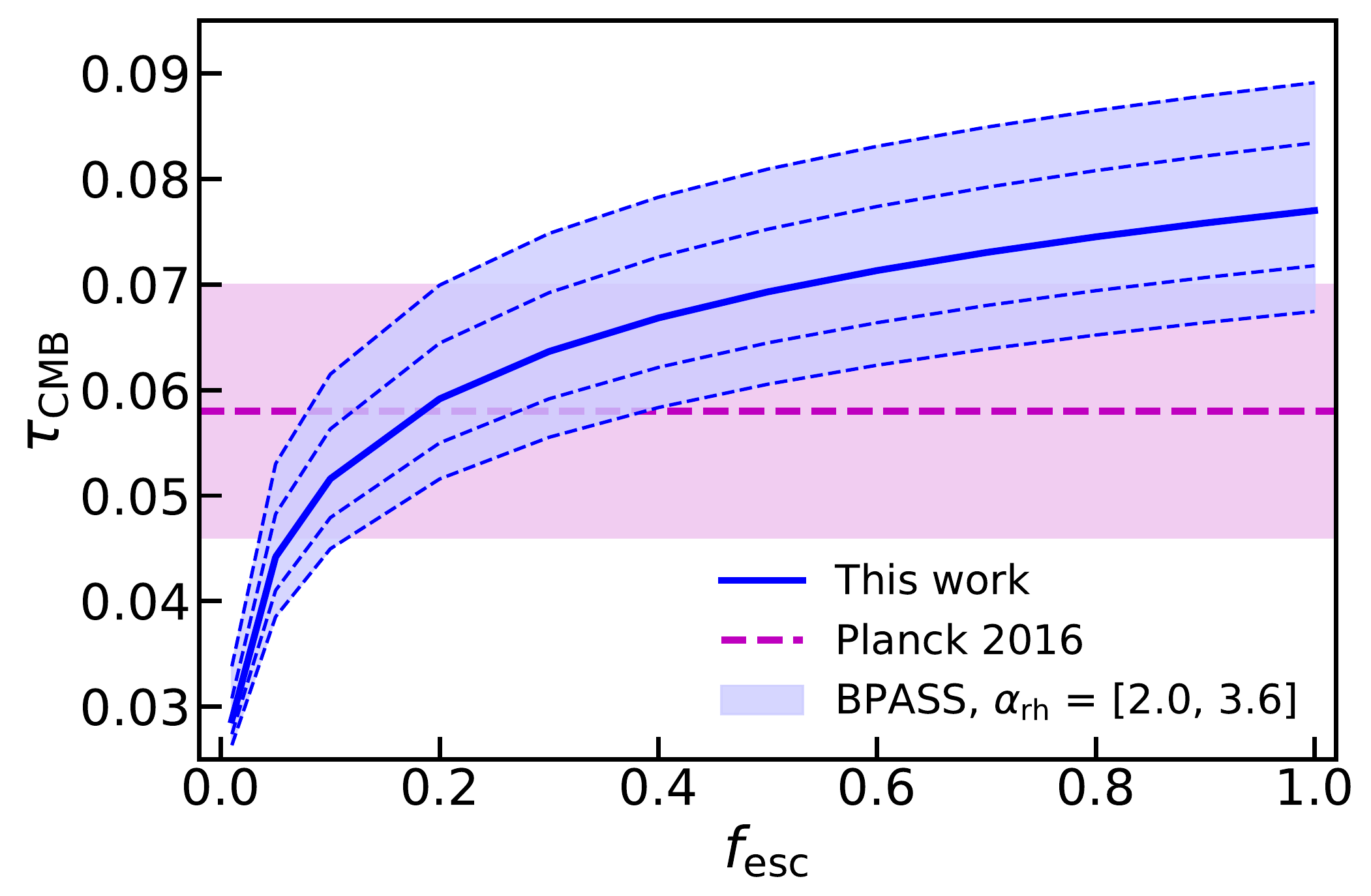}
    \caption{Thomson scattering optical depth of the CMB, \taucmb, as a function of assumed \fesc\ and $\alpha_\text{rh}$ in our model. Predictions from our reference model are shown by the blue solid line. The blue dashed lines show the alternative predictions made with $\alpha_\text{rh} = 2.0$ (upper bound for earlier reionization due to higher number density of galaxies), 2.4, 3.2, and 3.6 (lower bound for later reionization due to lower number density of galaxies). The measured value of $\tau_\text{CMB} = 0.058 \pm 0.012$ as reported by \citet{Planck2016a} is shown for comparison. This provides a reference showing the interplay between uncertainties or variation in the parameters \fesc\ and $\alpha_\text{rh} = 2.0$. }
    \label{fig:unified_tau}

\end{figure}

\begin{figure}
    \includegraphics[width=\columnwidth]{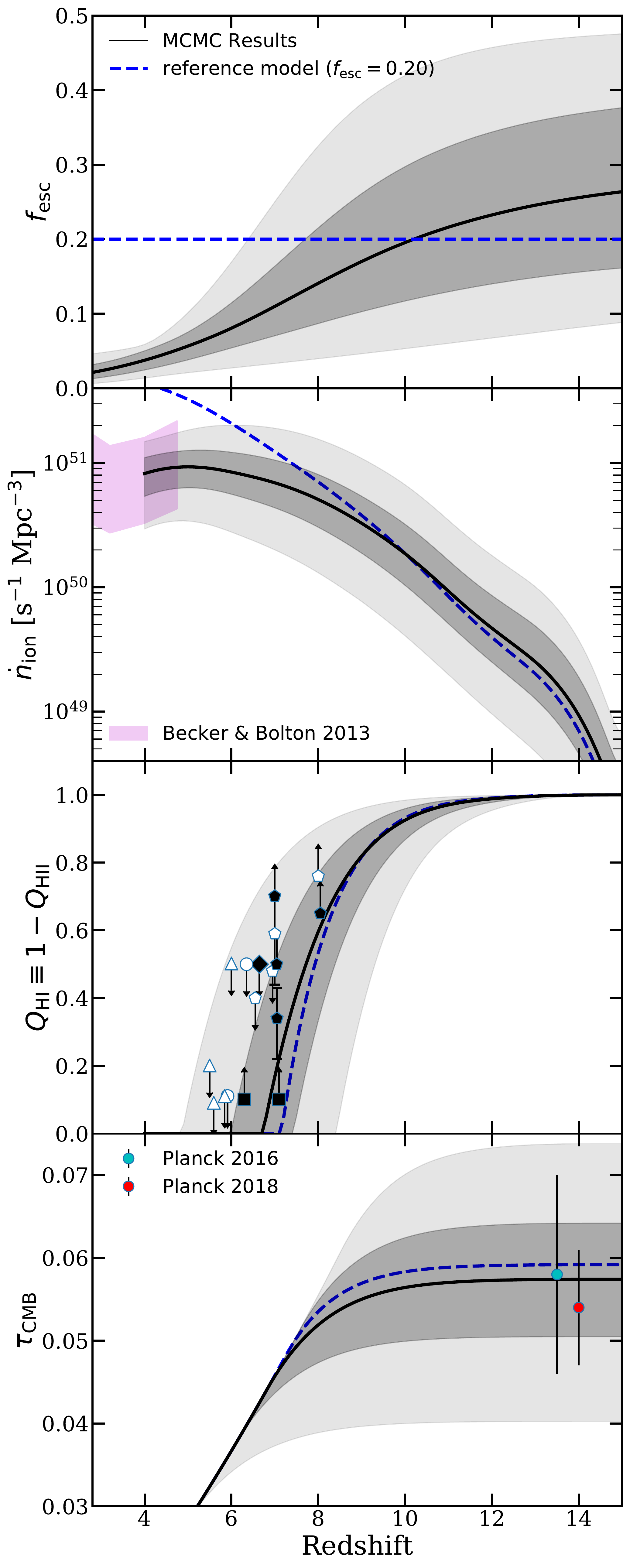}
    \caption{From top to bottom, we show the predicted redshift evolution for \fesc, \nion, \qhi, and \taucmb\ using the \citetalias{Becker2013} and  \citet{Planck2016a} \taucmb\ observations as constraints in our MCMC analysis. The shaded areas denote the 68\% (dark grey) and 95\% (light grey) confidence regions. We also show our reference model with a constant $f_\text{esc} = 0.20$ for comparison (blue dashed). This figure shows that a moderate effective evolution of \fesc\ with redshift can comfortably accommodate all of the observational constraints. }
    \label{fig:MCMC_results_all}
\end{figure}

Fig.~\ref{fig:unified_tau_z} shows \taucmb\ as a function of redshift for our reference model, and for the model variants $f_\text{esc}=[0.05, 0.80]$ and $\alpha_\text{rh}=[2.0, 3.6]$. 
We show recent measurements reported by the Planck Collaboration \citeyear{Planck2014, Planck2016a, Planck2018} and WMAP-9 \citep{Hinshaw2013}.
The latest observational constraints together favour both a later conclusion of reionization and a less rapidly evolving \nion, which ease both the need for high emissivity at high redshifts and rapid decrease of \nion\ toward $z \lesssim5$. 
In Fig.~\ref{fig:unified_tau}, we show the integrated \taucmb\ as a function of both \fesc\ and $\alpha_\text{rh}$.
This  shows that \taucmb\ is very sensitive to the LyC escape fraction for $f_\text{esc}\lesssim0.3$, but its dependence on \fesc\ becomes much flatter above this value. 
For $f_\text{esc}\gtrsim0.3$, the predicted optical depth is more sensitive to the abundance of faint galaxies rather than the LyC \fesc.
Note that \taucmb\ is an integrated quantity that compresses the reionization history into a single metric.
However, it is degenerately affected by both the conclusion of the phase transition and its progression. 
For instance, an extremely slow reionization progression or a rapid, late reionization can both result in a lower measured \taucmb\ value.

These results show that models with non-evolving \fesc\ and $\alpha_\text{rh}$ are unable to reproduce a reionization history that simultaneously matches all three sets of constraints, which is consistent with results from previous studies \citep[e.g.][]{Kuhlen2012, Anderson2017}.

\subsection{Constraining a redshift dependent escape fraction with MCMC}
\label{sec:variable}

Results from \S\ref{sec:constant} quantified the sensitivity of model outputs to a range of fixed values of \fesc\ and $\alpha_\text{rh}$.
In this section, we allow \fesc\ to evolve as a function of redshift (see eqn.\ref{eqn:fesc}) and employed a Markov chain Monte Carlo (MCMC) method to find the optimal configuration that is needed to satisfy the current observational constraints.
We employ the python MCMC tool \texttt{emcee}\footnote{\url{http://dfm.io/emcee}, v2.2.1} by \citet{Foreman-Mackey2013} to survey the four dimensional parameter space, including $f_\text{esc,max}$, $f_\text{esc,0}$, $k_0$, and $\alpha_\text{rh}$.
In the context of cosmic reionization studied here, we consider the many other free parameters in the galaxy formation model as being collectively constrained either by calibration or by the cross-checks with observations between $z=4$ to 10 in previous works.
In this exercise, $\alpha_\text{rh}$ can take any value within the range [2.0, 3.6], where \nion\ is pre-calculated for fixed values of $\alpha_\text{rh}=2.0$, 2.4, 2.8, 3.2, and 3.6, and then interpolated using the \texttt{scipy.interpolate} tool.
By varying $\alpha_\text{rh}$ within the SAM, we have included a number of associated features under its influences, including the faint-end slopes and flattening of the UV LFs, and the subtle boost of ionizing photon production due to the slight increase in burstiness triggered by SN feedback (see section 3.4 and associated discussion in \citetalias{Yung2020}).

The MCMC framework is set up with 56 walkers, each of which performs a chain of 120,000 steps with the first 200 regarded as burn-in and discarded.
Each of these walkers is initialized with a Gaussian distribution, with a chosen peak and half width distribution. 
The parameters that went into the set-up can be found in Table \ref{table:mcmcparam}.
We assumed a flat prior for all four of our free parameters.
For a randomly drawn prior that falls outside the boundary of the flat prior, a new set of parameters are drawn.

\begin{table}
    \centering
    \caption{Summary for the MCMC parameters, flat prior constraints, and posterior with 68\% confidence region.}
    \label{table:mcmcparam}
    \begin{tabular}{ l c c c c }
        \hline
        & Initiation & $\sigma$ & constraints & posterior \\
        \hline
        $f_\text{esc,0}$ & 0.036 & 0.00005 & [0.012, 0.060] & $0.0381^{+0.0148}_{-0.0159}$ \\ [0.1in]
        $f_\text{esc,max}$ & 0.350 & 0.0005 & [0.100, 0.500] & $0.2985^{+0.1357}_{-0.1328}$ \\ [0.1in]
        $k_0$ & 0.50 &  0.005 & [0.10, 0.90] & $0.523^{+0.255}_{-0.269}$ \\ [0.1in]
        $\alpha_\text{rh}$ & 2.80 & 0.05 & [2.0, 3.6] & $2.784^{+0.525}_{-0.511}$ \\
        \hline
    \end{tabular}
\end{table}

\begin{figure}
    \includegraphics[width=\columnwidth]{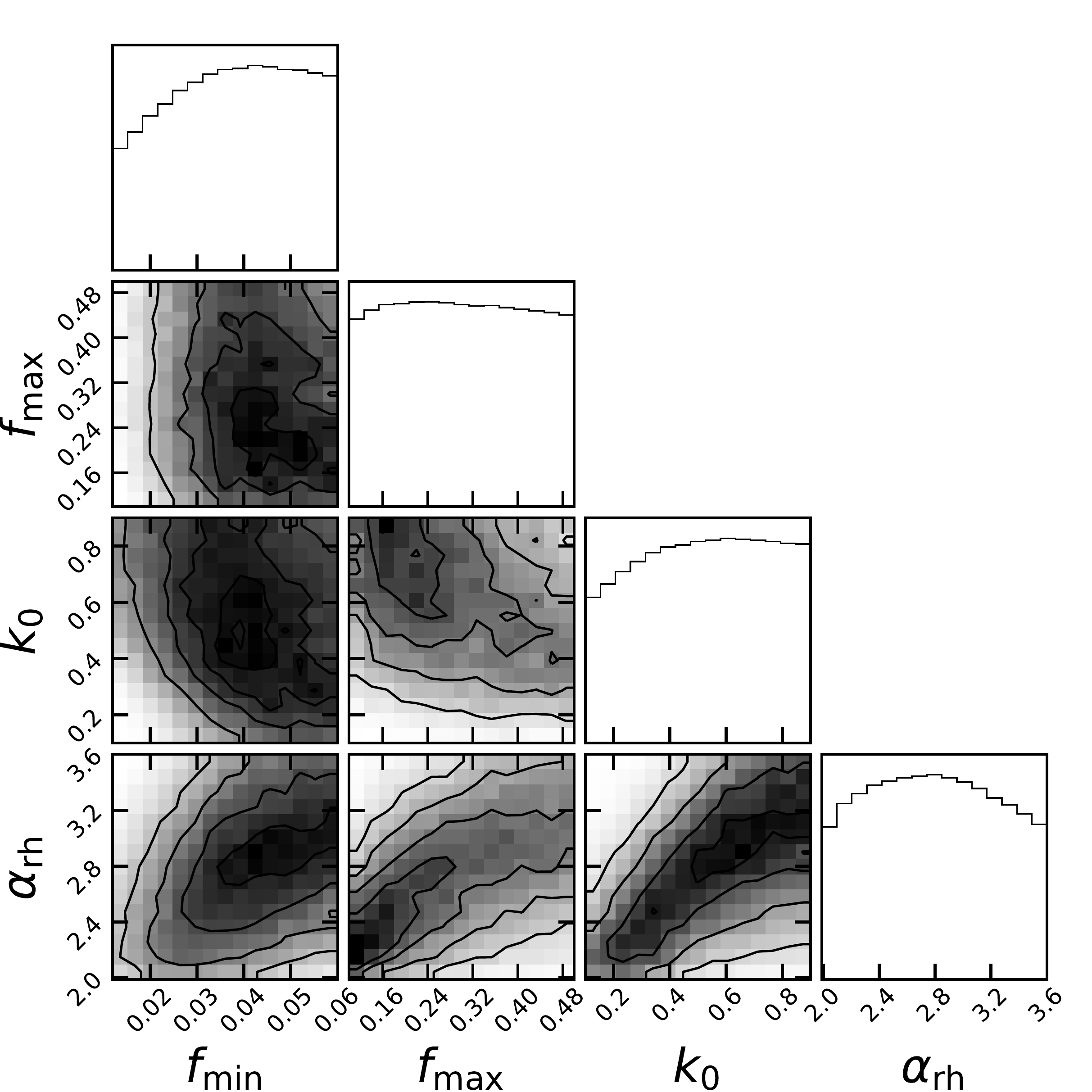}
    \caption{Distributions of the parameter posterior distributions from the MCMC. The parameter $\alpha_\text{rh}$ is fairly well constrained, as is the low-redshift (here $z=4$) value of \fesc\ ($f_\text{esc,0}$). The rate of evolution of \fesc\ ($k_0$) and its asymptotic value ($f_\text{esc,max}$) are not well constrained. The strong covariance between $\alpha_\text{rh}$ and \fesc\ is apparent.}
    \label{fig:MCMC_parameters}
\end{figure}

The set of observational constraints used in the MCMC are the \Lya\ forest constraints on \nion\ from \citetalias{Becker2013} and the \taucmb\ from \citet{Planck2016a}, which are weighted equally in the likelihood function.
Note that the large collection of IGM neutral fraction estimates are shown for comparison but are not used as constraints in the MCMC.
The median and the 68\% and 95\% confidence region of our posteriors for the predicted \fesc, \nion, \qhi, and \taucmb\ are summarized in Fig.~\ref{fig:MCMC_results_all}, where the posterior distributions are shown in Fig.~\ref{fig:MCMC_parameters}.

Our results favour a drop in escape fraction at $z\lesssim7$, leading to a turn-over in the ionizing emissivity. The parameters $k_0$ and $\alpha_\text{rh}$ are strongly covariant, and are only weakly constrained by \taucmb.
It is encouraging that the range of predicted reionization histories are in broad agreement with the \qhi\ constraints.
This is non-trivial as it depends on other model components that are not being actively `tuned' here, such as which galaxies are contributing to reionization.
We note that the latest \citealt{Planck2018} result would favour an even milder evolution of \fesc\ and a slightly lower $f_\text{esc,max}$. Adopting the Planck2018 value of \taucmb\ will only mildly change the results and conclusions of this work.

\subsection{Comparison with other recent models}
\label{sec:model_compare}

\begin{figure}
    \includegraphics[width=\columnwidth]{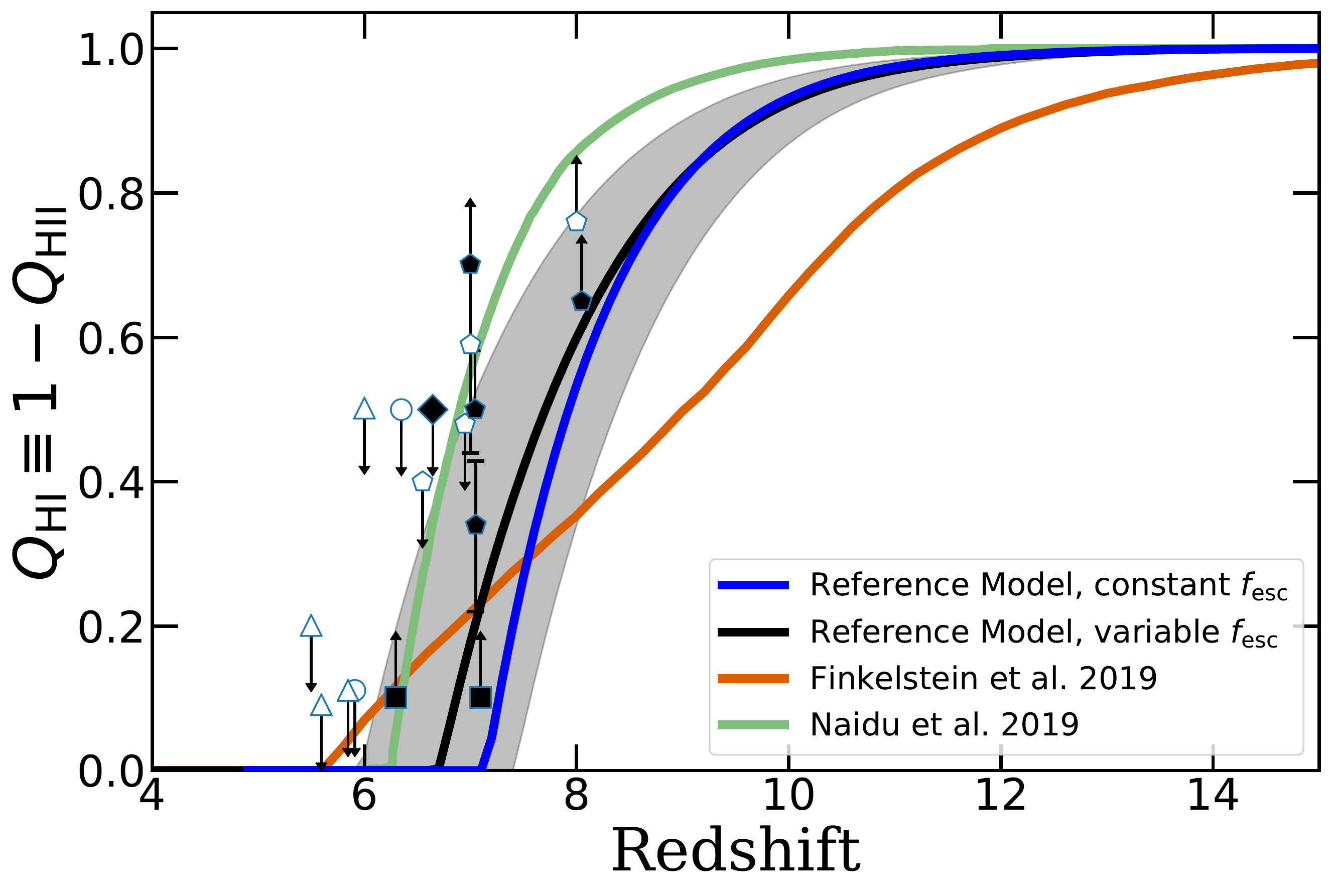}
    \caption{The redshift evolution of the IGM neutral fraction predicted by our reference model with a constant $20\%$ (blue) escape fraction, and our fiducial model with an evolving (black line and shaded regions) escape fraction, compared to predictions from the recent models of \citet{Finkelstein2019} (orange) and \citet{Naidu2020} (green). A compilation of observational constraints is shown by the symbols (see Fig. \ref{fig:unified_qhii} for a legend). This illustrates the broad range of reionization histories implied by several of the most recent modeling papers.}
    \label{fig:unified_qhii_model}
\end{figure}

In this section, we compare the predicted reionization history in our reference model to recent studies by \citet{Finkelstein2019} and \citet{Naidu2020}.
In fig.~\ref{fig:unified_qhii_model}, we compare results from our reference model with a constant $f_\text{esc} = 20\%$ and with the evolving $f_\text{esc}$ found in section \ref{sec:variable} to the results from \citeauthor{Finkelstein2019} and Model I (constant \fesc) from \citeauthor{Naidu2020}.
Although these models are fairly similar as they adopted a similar approach to modelling reionization, we note that these works adopted very different approaches to model galaxy populations and their evolution.

\citeauthor{Finkelstein2019} use UV LFs from observations with faint-end slopes extrapolated below the current detection limit. 
These UV LFs are truncated at halo masses corresponding to photoionization squelching and atomic cooling, which are obtained via abundance matching.
The many moving parts in the model, including the LyC escape fraction, halo truncation mass, the evolution of \xion, and the contribution of AGN, are optimized to fit a set of observational constraints using a MCMC machinery.
\citeauthor{Finkelstein2019} were particularly interested in exploring models that could satisfy all constraints on reionization while adopting a low ionizing escape fraction ($\lesssim 5$\% throughout the EoR). They adopted a halo-mass dependent parametrization of  \fesc\ based on hydrodynamic simulations from \citet{Paardekooper2015}, and paramterized \xion\ in terms of redshift and galaxy luminosity.

On the other hand, \citeauthor{Naidu2020} adopted galaxies from the \citet{Tacchella2018} empirical model (see comparison with our predictions in \citetalias{Yung2019a}) and 
estimated the production efficiency of ionizing photons using synthetic SEDs generated from the Flexible Stellar Population Synthesis \citep[FSPS;][]{Conroy2009, Conroy2010} and MESA Isochrones and Stellar Tracks \citep[MIST;][]{Choi2017a} for individual galaxies.
They have explored a model with fixed \fesc\ and one that scales as a function of $\Sigma_\text{SFR}$, and a range of truncation values $M_\text{UV}$.

\citeauthor{Finkelstein2019} found that in order for models with such universally low escape fractions to be viable, a rather high and rapidly evolving \xion\ is required. The range of \xion\ values are similar to, or even above, the observed values from \citet{Bouwens2016a}. However, \citetalias{Yung2020}, \citet{Wilkins2016a}, and \citet{Ceverino2019} have shown that such high values of \xion\ and such strong evolution are not `naturally' predicted in current self-consistent galaxy formation models. As we can see in fig.~\ref{fig:unified_qhii_model}, the \citeauthor{Finkelstein2019} model (in which reionization is heavily dominated by low-mass galaxies) predicts an early start to reionization and a more gradual evolution for \qhi. The \citeauthor{Naidu2020} model (in which massive galaxies play a more important role) predicts later and more rapid reionization. Curiously, our model lies somewhere in between, although it also predicts a fairly rapid transition in \qhi.

This comparison illustrates that there is still significant uncertainty in which galaxies dominate the reionization of the Universe and the details of how reionization progressed.  Future observations \textit{JWST} and other facilities will provide direct constraints on the source populations (as we explore further in the next section). Furthermore, as galaxies of different masses cluster very differently in space, these models would also have very different implications for the topology of reionization, which will eventually be probed with 21-cm intensity mapping experiments.

\subsection{Which galaxies reionized the Universe and will JWST see them?}
\label{sec:jwst_obs}

\begin{figure}
    \includegraphics[width=\columnwidth]{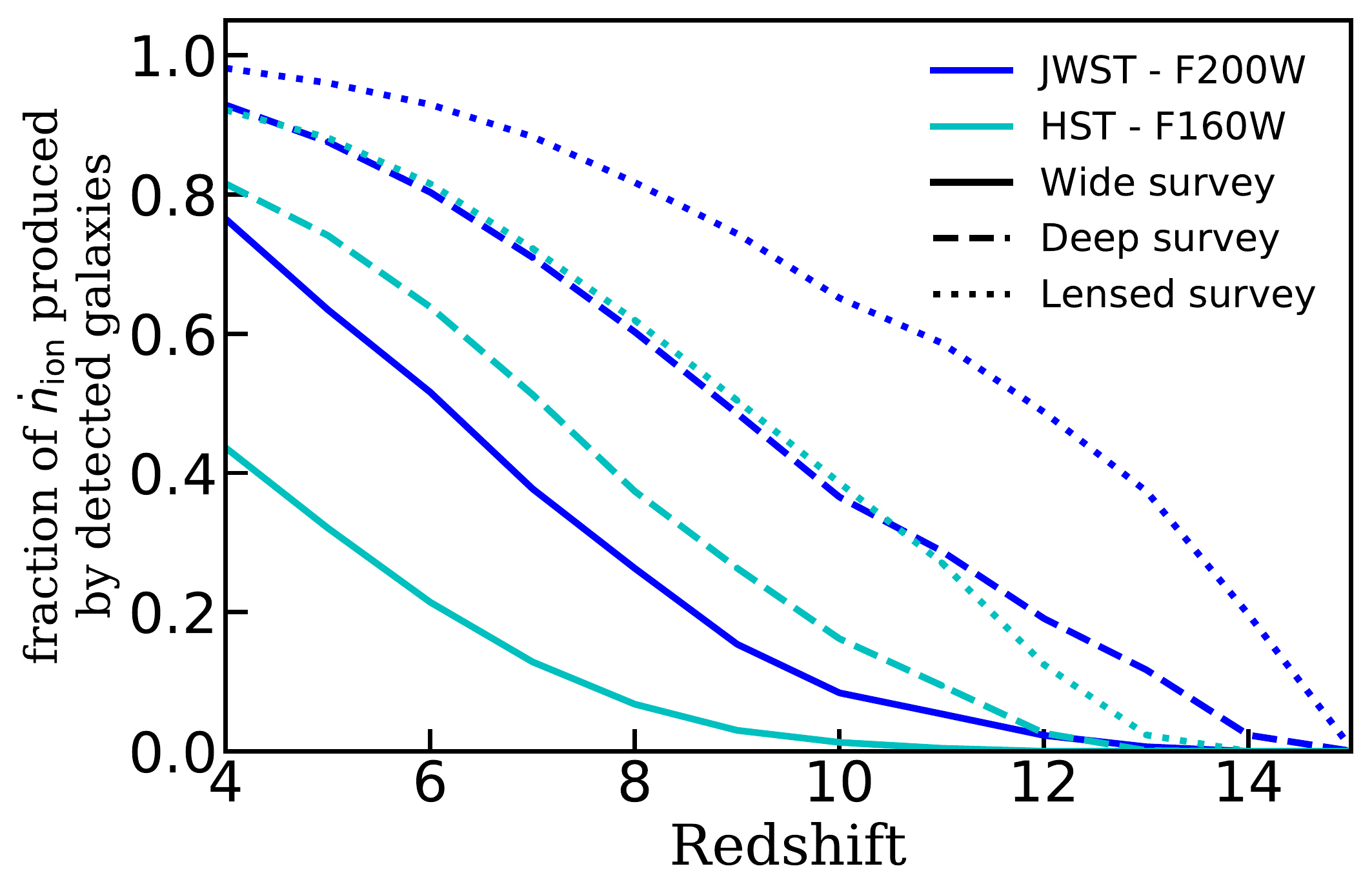}
    \caption{Predictions from our reference model for the fraction of ionizing photons produced by galaxies that are expected to be detected in various types of surveys, including wide (solid), deep (dashed), and lensed (dotted) surveys with \textit{JWST} (blue) and \textit{HST} (cyan) between $z=4$ and 15. Survey areas and detection limits assumed for these calculations are detailed in the text. These predictions reflect the production rate and do not account for the escape fraction of ionizing photons possibly varying across galaxies with different luminosities. }
    \label{fig:JWST_nionfrac} 
\end{figure}

In this section, we take advantage of the completeness of our predictions both in mass and redshift to estimate the contributions of ionizing photons from galaxies of different intrinsic luminosities, and estimate what fraction of the ionizing photon budget will be contributed by galaxies that are anticipated to be observed in future \textit{JWST} surveys. We show predictions for our reference model. As the dependence of \fesc\ on galaxy luminosity is very uncertain, and not yet included in our modelling, we only provide predictions here for the fraction of ionizing photons \emph{produced} and do not try to estimate the fraction that escape to the IGM. Recent simulation works have shown that \fesc\ may inversely scale with $M_\text{h}$ in a fairly loose way \citep[e.g.][]{Paardekooper2015}, and therefore the actual contribution from massive/luminous galaxies to the overall ionizing photon budget might be smaller than what is presented here.
However, our models do incorporate a mass and redshift dependent \xion\ based on our self-consistent modelling. These calculations do not account for field-to-field variance nor the survey area, where rare, massive objects may be missing from the small survey area of deep surveys.

In fig. \ref{fig:JWST_nionfrac}, we show the fraction of \nion\ produced by galaxies above the detection limits of hypothetical \textit{JWST} wide, deep, and lensed surveys with detection limits of $m_\text{F200W} = 28.6$, 31.5, and 34.0, respectively\footnote{The F200W filter on the NIRCAM instrument}. See Table 6 in \citetalias{Yung2019} and Table 1 in \citetalias{Yung2019a} for detailed configurations of these hypothetical surveys. For comparison, we also show results for legacy \textit{HST} surveys, where we adopted detection limits for the F160W filter $m_\text{F160W} = 26.8$, 29.5, and 31.5 for wide, deep, and lensed surveys, respectively, with configurations similar to the CANDELS and Hubble Frontier Fields surveys. At $z\sim 7$--8, where we predict the Universe to be about 50\% reionized by volume, \textit{JWST} will be able to detect the sources of 60-70\% of the reionizing photons in a deep survey. This fraction increases to  $\sim 90\%$ for an ultra-deep lensed survey, however, interpreting lensed observations and estimating the survey completeness may be more challenging.

\begin{figure}
    \includegraphics[width=\columnwidth]{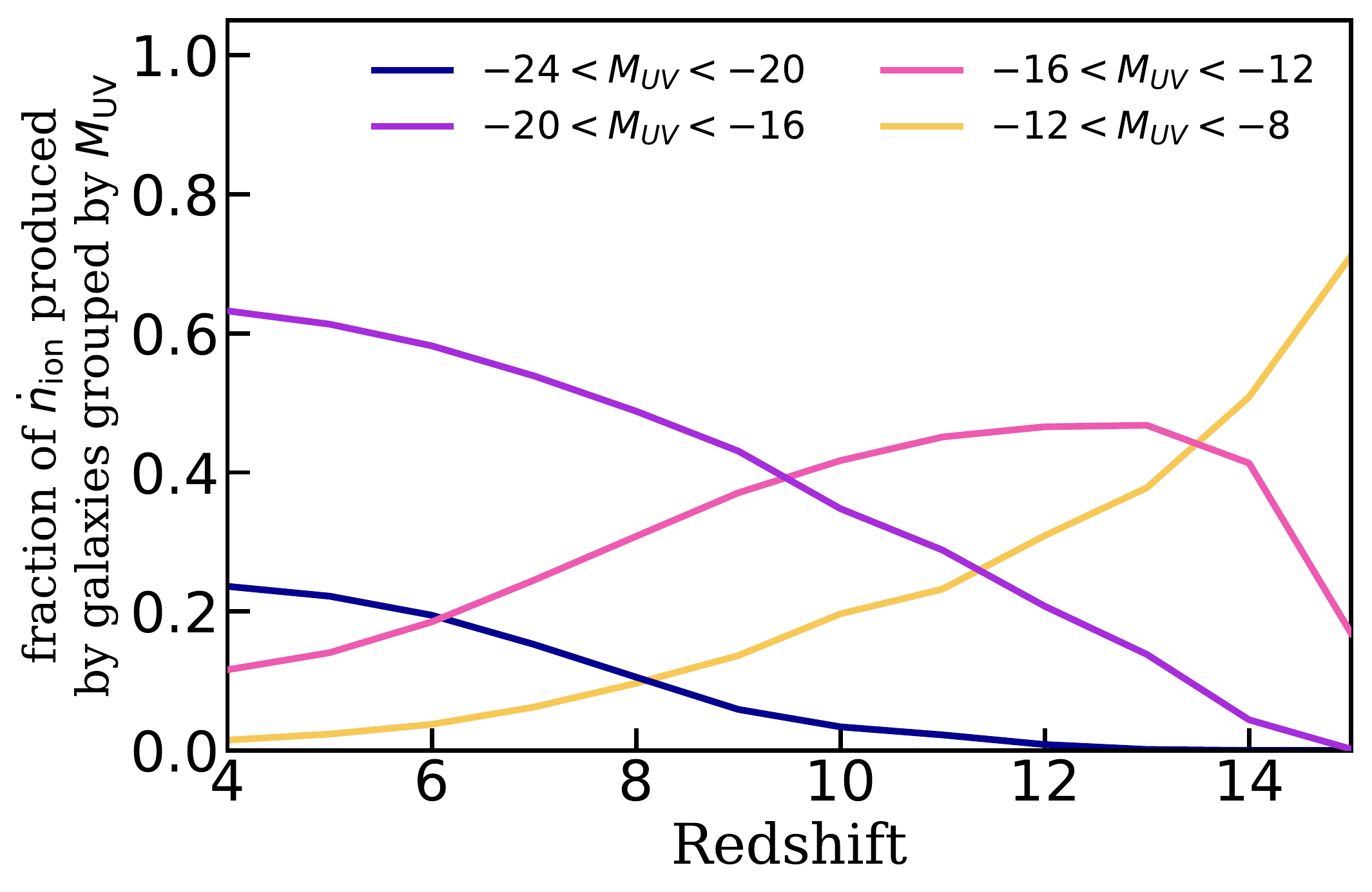}
    \caption{Predictions from our reference model for the fraction of ionizing photons produced by galaxies grouped by rest-frame dust-attenuated $M_\text{UV}$ between $z = 4$ and 15. These predictions reflect the production rate and do not account for the escape fraction of ionizing photons possibly varying across galaxies with different luminosities. Galaxies outside the range of $M_\text{UV}$ shown contribute $< 1\%$ of ionizing photons at all times.}
    \label{fig:grouped_nionfrac} 
\end{figure}

In a similar experiment, we break down the galaxy populations by rest-frame \emph{intrinsic} UV magnitude (not accounting for the effect of dust attenuation) into the following groups: $-24 < M_\text{UV} < -20$, $-20 < M_\text{UV} < -16$, $-16 < M_\text{UV} < -12$, to the faintest $-12 < M_\text{UV} < -8$.
In fig. \ref{fig:grouped_nionfrac}, we compare the fraction of \nion\ contributed by galaxies from each of these groups from $z=15$ to 4.
Galaxies beyond this range combined produce $< 1\%$ of ionizing photons across all redshifts, and are omitted here. Similar to results presented in the previous figure, we assume that \fesc\ does not depend on galaxy properties, which may significantly effect the predictions shown here. We find that ultra-faint galaxies ($-12 < M_\text{UV} < -8$) dominate at the highest redshifts ($z \gtrsim 13$), with a slightly brighter population $-16 < M_\text{UV} < -12$ dominating over the redshift range $10 \lesssim z \lesssim 13$. At lower redshift $z \lesssim 10$, galaxies in the intermediate luminosity range $-20 < M_\text{UV} < -16$ dominate.

\begin{figure}
    \includegraphics[width=\columnwidth]{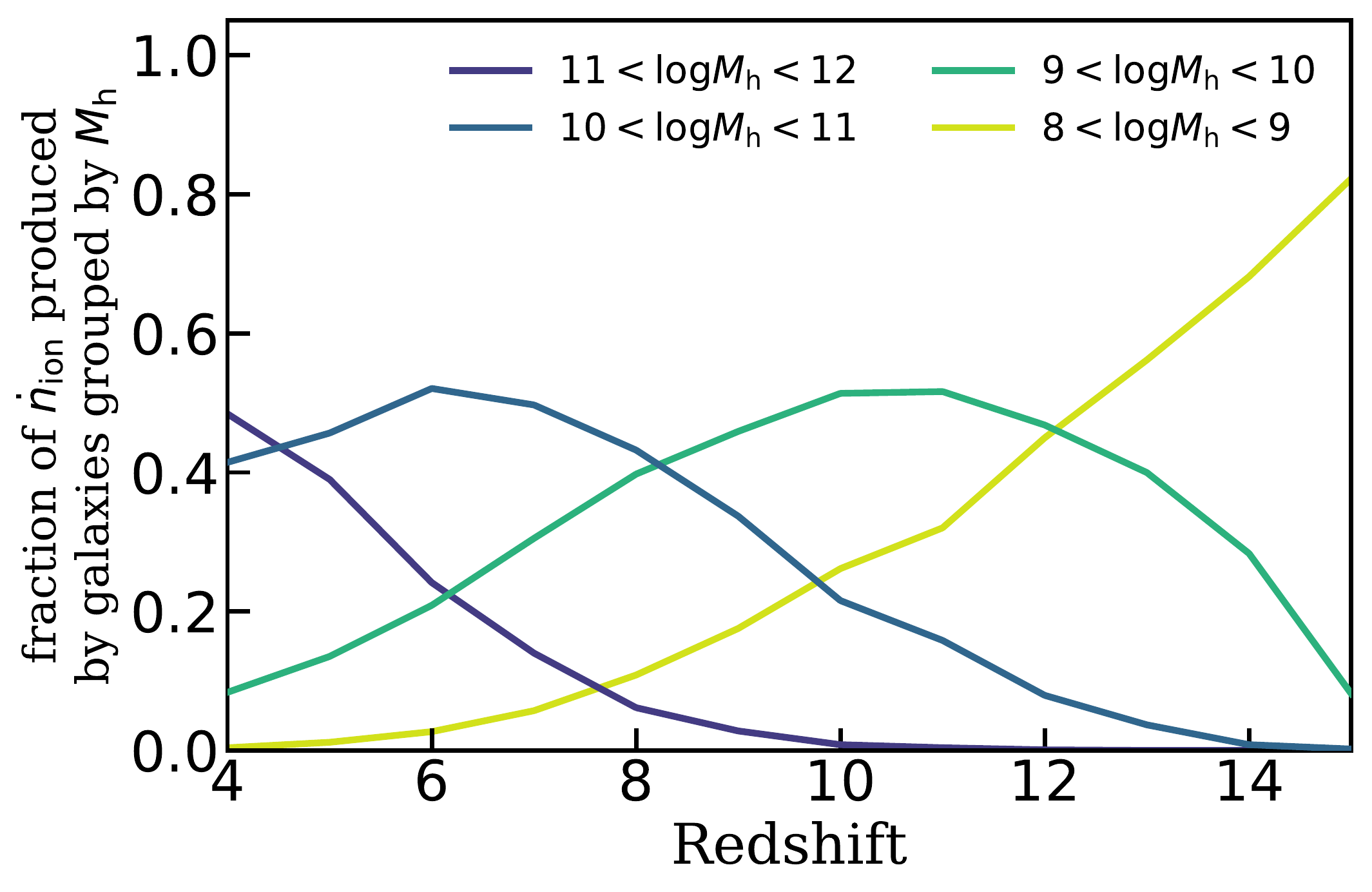}
    \caption{Predictions from our reference model for the fraction of ionizing photons produced by galaxies grouped by host halo mass $M_\text{h}$ between $z = 4$ and 15. These predictions reflect the production rate and do not account for the escape fraction of ionizing photons possibly varying across halo mass. The contribution of ionizing photons that originates in halos outside the range shown is insignificant.}
    \label{fig:grouped_halomass} 
\end{figure}

Similarly, in fig.~\ref{fig:grouped_halomass} we break down the contribution of ionizing photons by the host halo masses of galaxies. These are based on the predictions from our reference model configurations, and are quite sensitive to the details of how galaxies populate halos, which as we have shown depends on the details of the stellar feedback parameters and other physical processes. 
We find that contributions from halos outside the range shown here are insignificant. 
This result is also useful for estimating the `completeness' of the predicted ionizing emissivity from studies with limited mass resolution. 
In \citetalias{Yung2019} (see section 2.2 and fig.~3), we explored the impact on star formation from a photoionizing background using a redshift-dependent characteristic mass approach as described by \citet{Okamoto2008} and found nearly no impact on the galaxy populations at the range of redshift and halo mass relevant to our study. However, high-resolution hydrodynamic simulations have shown that the presence of such a background may have affected the low-mass, `photosensitive' halos of $\log(M_\text{h}) \lesssim 9$ \citep{Finlator2013}. Accounting for this effect may reduce the contribution from low-mass halos near the beginning of the EoR relative to our predictions.

\section{Discussion}
\label{sec:discussion}

In this section, we discuss some caveats and uncertainties in our
modelling pipeline, and present an outlook for future observations
with \textit{JWST} and beyond.

\subsection{Galaxies forming at extreme redshifts and their role in cosmological events}

In this series of papers, we have explored the interplay between galaxy formation physics and the cosmological-scale phase transition of hydrogen reionization. In particular, we investigated whether models with physical recipes and parameters that have been calibrated to match lower redshift observations ($z\sim 0$) are consistent with a broad suite of observations at extremely high redshifts ($z\gtrsim6$). A significant finding of this work is that \emph{these locally calibrated models are consistent (within the uncertainties) with all currently available observations at $z\gtrsim6$, including direct observations of galaxies, and indirect probes of the reionization history from observations of the IGM and CMB.} This has two important implications: 1) It seems that the physical processes regulating star formation and stellar feedback do not operate in a vastly different manner at extremely high redshift. Given our lack of detailed understanding of how these processes work even in the local Universe, this is far from a trivial conclusion. 2) Contrary to some previous suggestions in the literature, the current suite of observations do not \emph{require} an additional `exotic' population of reionizing sources (other than galaxies, e.g. mini-quasars, Pop III stars, self-annihilating dark matter, etc.) in the early Universe. The remaining uncertainties on several components of our modelling mean that we do not \emph{rule out} the existence of such sources at some level -- but they are not required to satisfy existing constraints. 

The overall ionizing photon budget available during the EoR is degenerately affected by physical processes that operate over a vast range of scales. As illustrated in fig.~\ref{fig:unified_nion}, each of these seemingly degenerate components can evolve differently and be constrained independently. 
In \citetalias{Yung2019} and \citetalias{Yung2019a}, we provided physically motivated predictions for the evolution of the number density of galaxies at high redshift, which will be further constrained with future galaxy surveys; and in \citetalias{Yung2020}, we predicted the evolution and distribution of \xion\ for the same set of galaxies, which may also be constrained with future observations as discussed in \citetalias{Yung2020}.
We have explicitly broken down the contribution to the ionizing photon budget as a function of redshift from galaxies with different observed frame and rest-frame luminosity, and different halo mass. This is again a non-trivial calculation, as the intrinsic production efficiency of ionizing photons depends on a combination of factors such as stellar population age and metallicity in addition to the number density of galaxies with different luminosities. These effects are self-consistently included in our models.

We took a fully semi-analytic approach to assemble our modelling pipeline, including the construction of merger trees, formation and evolution of galaxies, and the progression of cosmic reionization. 
In practice, this modelling pipeline serves as a low-computational-cost platform for examining galaxy formation and cosmic reionization constraints from various tracers. 
As shown in fig.~\ref{fig:MCMC_results_all}, the set of IGM neutral fraction constraints seem to collectively favour a relatively rapid decline of neutral hydrogen around $z \sim 6$--7. However, such a reionization history would yield a \taucmb\ near the lower-bound of the reported uncertainties of the latest measurements.
Although these constraints are in mild tension, our model is in agreement with all constraints within the 68\% confidence regions of the MCMC posterior.
Adopting the even lower \taucmb\ measurement reported by the \citeauthor{Planck2018} in \citeyear{Planck2018}  would further ease this tension and yield a slightly milder evolution for \fesc\ and a slightly more gradual reionization history. 

Another novel aspect of our work is the rigorous statistical exploration of the degeneracy in the physical parameter controlling the impact of stellar feedback on low-mass halos ($\alpha_\text{rh}$) and the parametrized effective redshift evolution of \fesc. Larger values of $\alpha_\text{rh}$ result in stronger feedback, producing fewer low-mass galaxies, and require higher values of \fesc\ to produce the required budget of ionizing photons, and vice versa. 
Under the subset of observational constraints included in the MCMC, 
it is encouraging that the median of the posterior of $\alpha_\text{rh}$ is very much in agreement with the value that is required to reproduce $z\sim0$ observations. 
Similarly, the required redshift evolution in \fesc\ is not extreme, and the values at the lower end of our explored redshift range ($z\sim 4$) where there are some observational constraints are reasonable. 

Past studies predicted diverging scenarios for the final stage of reionization depending on the assumptions and observational constraints employed by these models. Such process could be rather extended when dominated by low-mass galaxies \citep[e.g.][]{Finkelstein2019}, or conversely very rapid when dominated by massive galaxies (e.g. \citealt{Mason2018} and \citealt{Naidu2020}), for which the rapid end to reionization is motivated by \Lya\ emitter constraints. 
The results presented in this work depict a relatively early onset of reionization compared to \citeauthor{Naidu2020} due to the early contributions from low-mass galaxies, but lag behind \citeauthor{Finkelstein2019} because of the lower predicted \xion\ and \fesc. 
However, it is very intriguing to see that our model also predicted a very rapid end to reionization when \Lya\ emitters are not explicitly used to constrain our model.
Fig. \ref{fig:grouped_nionfrac} shows that the contribution of ionizing photons from more massive galaxies has grown rapidly and took over from their low-mass counterparts during the EoR, which provides a physical explanation to the rapid conclusion of reionization that is solely driven by galaxy formation physics rather than from observed EoR constraints. 

However, \citeauthor{Finkelstein2019} also showed that by letting all galaxies to have the same escape fraction, galaxies with $M_\text{UV} \lesssim -16$ would have dominated the ionizing photon budget.
Given that the way \fesc\ is parametrized in this work, it is also possible that we have overestimated the contribution from massive galaxies, which could be a partial reason to the rapid end.
Therefore, the predicted rapid end to reionization can be one part backtracked to the predicted evolution of galaxy populations and their spectroscopic properties, and one part due to our parametrization of \fesc.

\subsection{Caveats, limitations, and uncertainties of the modelling framework}

The limitations and caveats regarding the galaxy formation model and the physical recipe for \Nion\ have been thoroughly discussed in previous works; we refer the reader to section 6.3 in \citetalias{Yung2019a} and section 4.3 in \citetalias{Yung2020}.
This discussion will be focused mainly on the topics related to the reionization pipeline presented in this work.

We note that even though the SAM is fairly successful at reproducing a wide variety of existing observational constraints, which we examined in detail in \citetalias{Yung2019} and \citetalias{Yung2019a}, both the physical properties and number density of the predicted galaxy populations at $z > 10$ are poorly constrained due to the lack of direct observations. They are subject to uncertainties in model components, such as feedback effects and SF relations, which are either untested or known to be inaccurate in extreme (e.g. metal-free) environments. There are also missing physical processes, such as the formation of Population III stars, that can potentially affect star formation activity in low-mass halos in the early universe. 

Therefore, we regard the predictions for $z=11$ to 15, including both the UV LFs and \Nion/$M_h$, to be more uncertain. We plan to explore the physics relevant to these extreme epochs in future works. In addition, the models will be tested more stringently as high-redshift observational constraints from \textit{JWST} and other instruments become available.

Furthermore, the EPS-based merger trees adopted in this work series have been compared trees extracted from numerical simulations and the results shown to be in good agreement. However, the EPS algorithm has never been tested over the full halo mass and redshift ranges that are explored in this work, as there is currently no publicly available relevant suite of dark matter only simulations. We plan on running and analysing this suite of N-body simulations, and developing and validating a new and improved fast merger tree algorithm, in Yung et al. (in prep). 

Our analytic reionization model does not account for density fluctuations and clustering of sources across the Universe, which numerical simulations have shown can lead to an inhomogeneous and `patchy' progression of reionization. Furthermore, our models do not self-consistently account for photo-ionization feedback (or `squelching'). Other works have shown \citep{Gnedin2014, Mutch2016} that photo-ionization affects only galaxies in extremely low mass halos, which we found have a negligible contribution to reionization. Furthermore, post-reionization IGM temperature fluctuation can also be used to constrain the reionization history and this has been explored analytically \citep {Furlanetto2009} and with fully coupled radiation-hydrodynamic simulations \citep{Wu2019}.

Some sources that could be potential contributors to the total ionizing photon budget are not accounted for in this work. These include Population III stars, X-ray binaries, and AGN. Previous work has shown that Pop III stars are unlikely to dominate the reionizing photon budget \citep[e.g.][]{Ricotti2004, Greif2006, Ahn2012, Paardekooper2013, Robertson2015}, but they could make reionization more patchy, and are presumably important for polluting early halos with metals, which can then provide the seeds for dust and molecular hydrogen formation. This process is a critical component in our models, which is currently treated in a simplified way by adopting a metallicity `floor' in all pristine halos. We further assume that significant cooling cannot occur in halos below the atomic cooling limit ($10^4$ K). While some cooling may occur at lower temperatures due to molecular hydrogen cooling or metal cooling, these are thought to be sub-dominant \citep{Yoshida2004, Maio2010, Johnson2013, Wise2014, Xu2016, Jaacks2018a}. The contribution of early accreting black holes to reionization remains very uncertain, and we plan to investigate this in upcoming work. In addition to directly contributing ionizing photons through their hard spectrum, semi-analytic calculations have shown that X-rays produced by AGN can boost \fesc\ \citep{Benson2013, Seiler2018}. However, hydrodynamic simulations have shown that this effect is not significant \citep{Trebitsch2018}. We also note that the contribution from AGN to the total ionizing photon budget can become fairly significant near the completion of \ion{H}{I} reionization \citep[e.g.][]{Dayal2020}, and our results matching the \citetalias{Becker2013} emissivity constraints may imply an over-prediction of the contribution from galaxies.

Another caveat related to the observational constraints is that the estimates of \taucmb\ are highly covariant with other cosmological parameters, and are derived assuming a simple instantaneous reionization model. As additional constraints on \qhi\ and the ionizing photon emissivity are obtained, and we gain a better understanding of the uncertainties on these measurements, these could be incorporated as additional constraints in a fitting procedure.

\subsection{Constraining galaxy formation during the EoR with \textit{JWST} and beyond}

With both deep- and lensed-field NIRCam surveys anticipated to reach unprecedented detection limits, the extremely sensitive \textit{JWST} is expected to directly detect and constrain the number density of faint galaxies up to $z\sim10$.
Furthermore, MIRI and NIRSpec will provide high-resolution spectroscopic follow-ups for the spectral features of these galaxies, which will put more robust constraints on \xion\ and \fesc. These measurements will allow us to further test and refine galaxy formation models and to understand the physics that shapes galaxy properties at ultra-high redshift. 

The coming decades promise great opportunities for further exploring the high-redshift Universe. The line-up of flagship instruments, include space-based \textit{Euclid} \citep{Racca2016} and \textit{Wide-Field Infrared Survey Telescope} \citep[\textit{WFIRST},][]{Spergel2015}, as well as the ground-based Large Synoptic Survey Telescope \citep[LSST,][]{LSST2017}. These facilities are capable of surveying large areas, which is complementary to the small field-of-view of \textit{JWST}. 
Furthermore, next generation facilities European Extremely Large Telescope \citep[ELT,][]{Gilmozzi2007}, Thirty Meter Telescope \citep[TMT,][]{Sanders2013}, and Giant Magellan Telescope \citep[GMT,][]{Johns2008} have the capability of doing spectroscopic follow-up on the expected large number of photometric detections.
The flexibility of our model allows it to be easily adapted to made predictions for these instruments, and facilitate physical interpretation for future multi-instrument surveys. 
In addition, the Atacama Large Millimeter Array (ALMA) has the capability of detecting dust continuum as well as fine structure lines such as [CII] and [OII] of $z>6$ galaxies.
With the extended modelling framework presented in \citep{Popping2019} coupled with our SAMs, we will also be able to make predictions for joint \textit{JWST}--ALMA multi-tracer surveys. 

Intensity mapping is a complementary approach that surveys large areas of the sky at relatively coarse angular resolution, potentially providing direct constraints on the conditions of the intergalactic hydrogen and indirect, collective constraints on high-redshift galaxy populations \citep{Visbal2010, Visbal2011, Kovetz2017}. 
Numerous intensity mapping experiments for \ion{H}{I}, CO, \ion{C}{II}, and \Lya\ are planned or underway, including BINGO \citep{Battye2013}, CHIME \citep{Bandura2014}, EXCLAIM \citep{Padmanabhan2018}, HERA \citep{DeBoer2017}, HIRAX \citep{Newburgh2016}, Tianlai \citep{Chen2012}, LOFAR \citep{Patil2017}, MeerKat \citep{Pourtsidou2016, Santos2017}, CONCERTO \citep*{Serra2016}, PAPER \citep{Parsons2010}, etc., which together pave the way to future large-scale multi-tracer intensity mapping surveys. 
These observations can also be cross-correlated with galaxy surveys for a comprehensive view of the interaction between galaxies and the cosmic environment. The modelling framework presented here can also provide a powerful tool for efficiently producing physically self-consistent, multi-tracer predictions for intensity mapping experiments (Yang et al. in prep).

Finally, improving radiative hydrodynamic simulations of early galaxy evolution \citep[e.g.][]{Finlator2018, Wu2019} will complement our approach by providing more physically motivated priors for our key physical parameters, and suggesting new parametrizations that connect quantities such as escape fraction to galaxy properties \citep[e.g.][]{Seiler2019} rather than redshift, as we have assumed here.  Our approach provides a framework to bridge these detailed self-consistent models with upcoming deep and wide surveys to optimally constrain the physics of early galaxy formation.

\section{Summary and Conclusions}
\label{sec:snc}

In this work, we constructed a physically motivated, source-driven
semi-analytic modelling pipeline that links galaxy formation to the
subsequent reionization history using an analytic model for reionization.
The galaxy formation model has been tested extensively and shown to match extremely well with observational constraints up to $z\sim10$ in previous works, and we extended these predictions up to $z\sim15$. We have calculated \Nion\ self-consistently, accounting for the stellar age and metallicity distribution of the stellar population in each galaxies using state-of-the-art SPS models. We presented predictions for the ionizing emissivity, IGM neutral fraction, and Thomson optical depth to CMB throughout the Epoch of Reionization, and compared these to a wide range of observational constraints.  In a controlled experiment, we isolated and quantified the effect of each of the major moving parts in the total ionizing photon budget. We also explored two different scenarios with a constant and a redshift-dependent \fesc, and determined the required conditions for the predicted galaxy populations to reionize the Universe in the time frame require by IGM and CMB constraints. We explored the covariance of different model components (including \fesc\ and the efficiency of stellar feedback) using MCMC.

We summarize our main conclusions below.

\begin{enumerate}
    \item Using a well-tested physical galaxy formation model, which was calibrated only to $z\sim0$ observations and has been shown to well-reproduce observed distributions from $z\sim4$--10, we provide predictions for rest-frame UV luminosity functions and ionizing photon production rate for galaxies up to $z=15$.

    \item Adopting a non-evolving escape fraction of $\sim20\%$, the galaxy population predicted by our model yields sufficient amounts of ionizing radiation to be consistent with constraints from the Thomson optical depth \taucmb. However, this model is in tension with low-redshift \Lya\ observations on the IGM neutral fraction and observational constraints on the ionizing emissivity at $2 \lesssim z \lesssim 6$.

    \item We performed a number of controlled experiments to explore the impacts on the reionization history of varying the three main model components that influence the total ionizing photon budget, including the abundance of low-mass galaxies, intrinsic ionizing photon production rate, and LyC escape fraction. We find that the uncertainty on estimates of the total LyC emissivity is dominated by uncertainties on \fesc, with the strength of stellar feedback being the second most important factor.

    \item  We used MCMC to explore the covariance in these two parameters (\fesc\ and $\alpha_\text{rh}$, which parametrizes the efficiency of stellar feedback in low-mass halos). We parametrized the population averaged \fesc\ as a function of redshift, and jointly constrained these parameters along with $\alpha_\text{rh}$ using constraints from \Lya\ forest observations and \taucmb\ measurements.  We found that a `population-averaged' escape fraction that mildly increases from $\sim4\%$ to $\sim29\%$ between $z\sim4$ to 15 satisfies both constraints.

    \item We presented predictions for the fraction of ionizing photons produced by galaxies of different rest-UV luminosity as a function of redshift, and for the fraction of the total ionizing photon budget sourced by galaxy populations that will be observable in upcoming surveys with \textit{JWST}. At $z\sim 7$--8, where we predict the Universe to be about 50\% reionized by volume, we predict that \textit{JWST} will be able to detect the sources of 60--70\% of the reionizing photons in a deep survey, and up to $\sim 90\%$ in an ultra-deep lensed survey.
\end{enumerate}

\section*{Acknowledgements}
The authors of this paper would like to thank Viviana Acquaviva, Ryan Brennan, Richard Ellis, Fangzhou Jiang, Daisy Leung, Adam Lidz, Dan Foreman-Mackey, Takashi Okamoto, Viraj Pandya, Aldo Rodr\'iguez-Puebla, David Spergel, Kung-Yi Su, Stephen Wilkins, and Eli Visbal for useful discussions. 
We also thank the anonymous referee for the constructive comments that improved this work.
AY and RSS thank the Downsbrough family for their generous support, and gratefully acknowledge funding from the Simons Foundation.
AV gratefully acknowledges funding from the USF Faculty Development Fund and Research Corporation.
This work is supported by HST grant HST-AR-13270.001-A.
We also thank the Co-I's on the original HST proposal that inspired this work, which eventually turned into this paper series.

\section*{Data availability}
The data underlying this article are available in the Data Product Portal hosted by the Flatiron Institute at \url{https://www.simonsfoundation.org/semi-analytic-forecasts-for-jwst/}.



\bibliographystyle{mnras}
\bibliography{library.bib}



\appendix
\section{HMF fitting parameters for the extended redshift range}
\label{appendix:a}
\setcounter{table}{0} \renewcommand{\thetable}{A\arabic{table}}

\begin{figure}
    \includegraphics[width=\columnwidth]{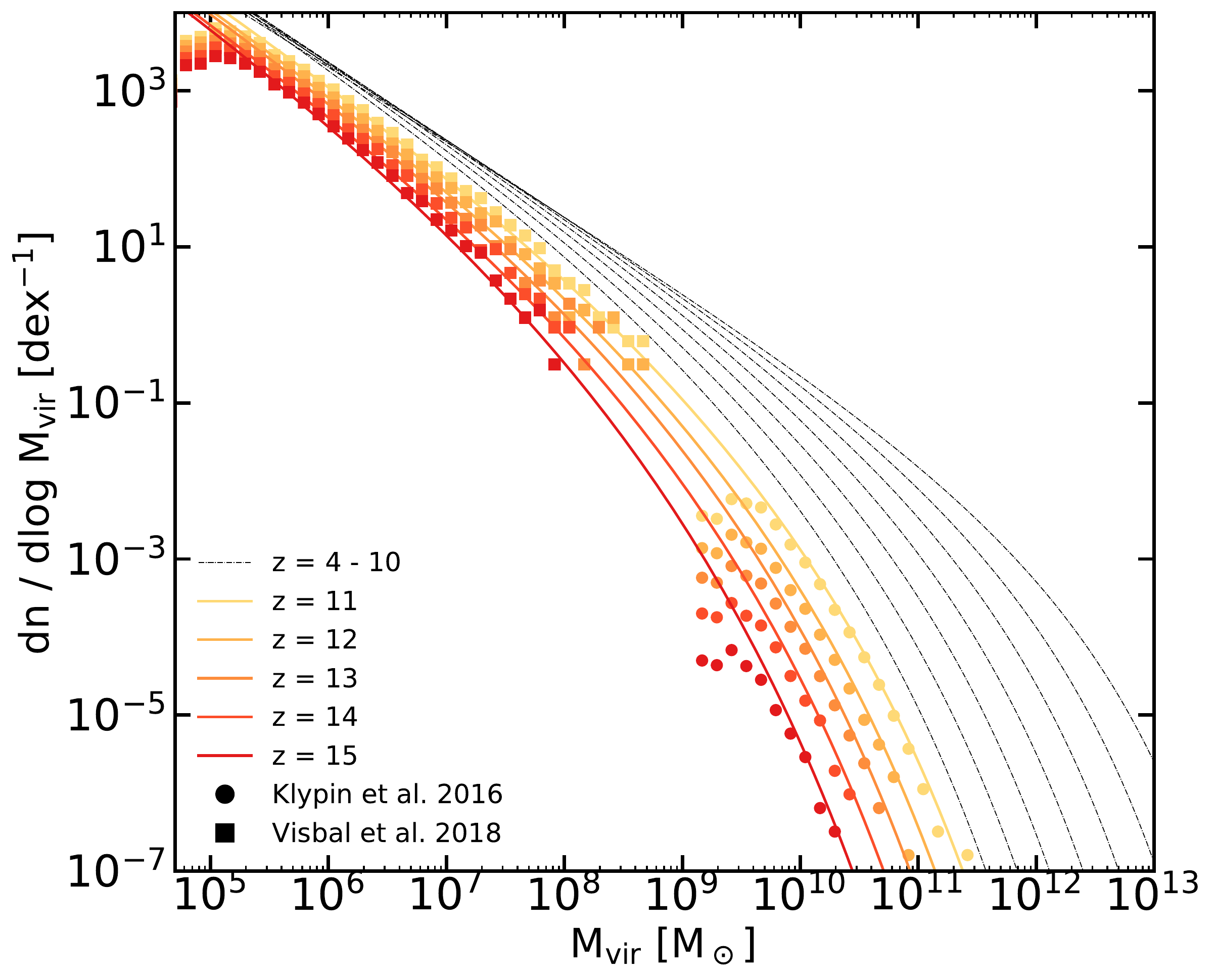}
    \caption{The coloured lines, from light to dark, show the HMF fitting functions adopted for the extended redshift range $z = 11$ -- 15. $n$-body simulation predictions from \citet{Klypin2016} and \citet{Visbal2018} are shown in matching colour for each redshift for comparison. The light grey dot-dashed lines show HMF fitting functions used for $z = 4$ -- 10 to guide the eye (see \citetalias{Yung2019} for detail).}
    \label{fig:HMF_ext}
\end{figure}

We adopted the HMF parametrization from \citet{Tinker2008} with parameters calibrated to the Bolshoi-Planck simulation from the MultiDark suite \citep{Rodriguez-Puebla2016, Klypin2016}. The comoving number density of halos of mass between $M_\text{vir} + d M_\text{vir}$ is given by
\begin{equation}
    \frac{dn_\text{h}}{dM_\text{vir}} = f(\sigma)\frac{\rho_m}{M^2_\text{vir}} \left| \frac{d \ln \sigma^{-1}}{d \ln M_\text{vir}} \right| \text{,}
\end{equation}
where $\rho_m$ is the critical matter density in the Universe, $\sigma$ is the amplitude of the perturbations, and $f(\sigma)$ is called the halo multiplicity function, which takes the form of
\begin{equation}
    f(\sigma) = A \left[ \left(\frac{\sigma}{b}\right)^{-a}\right]e^{-c/a^2} \text{,}
\end{equation}
where $A$, $a$, $b$, and $c$ are free parameters. In this work, as shown in fig. \ref{fig:HMF_ext}, we recalibrate these parameters to match the HMF constraints between $z = 11$ -- 15 from the Bolshoi-Planck simulation and from \citet{Visbal2018}. These parameters are presented in Table \ref{table:hmf_param}.

\begin{table}
    \centering
    \caption{Fitting parameters for $f(\sigma)$ parameters that produces the HMF at $z = 11$ -- 15 used throughout this work as shown in fig. \ref{fig:HMF_ext}.}
    \label{table:hmf_param}
    \begin{tabular}{ c c c c c }
        \hline
        $z$  & $A$ & $a$ & $b$ & $c$ \\
        \hline
        11 & 0.1668 & 0.9823 & 1.100 & 1.0938 \\
        12 & 0.1468 & 0.9823 & 1.000 & 1.0938 \\
        13 & 0.1468 & 0.9823 & 0.900 & 1.0938 \\
        14 & 0.1268 & 0.9823 & 0.750 & 1.0938 \\
        15 & 0.1268 & 0.5523 & 0.600 & 1.1238 \\
        \hline
    \end{tabular}
\end{table}

\section{Predicted UV Luminosity Functions}
\label{appendix:b}
\setcounter{table}{0} \renewcommand{\thetable}{B\arabic{table}}

Tabulated UV LFs from our fiducial model in the extended redshift range $z = 11$ -- 15 inducing dust attenuation are presented in Table \ref{table:UVLFs_ext}. See online data repository for other data presented in this work.

\begin{table}
    \centering
    \caption{Tabulated UV LFs at $z = 11$ -- 15 from our fiducial model \emph{with} dust attenuation as presented in fig. \ref{fig:UVLFs_everything} (solid lines).}
    \label{table:UVLFs_ext}
    \begin{tabular}{cccccc}
        \hline
        & \multicolumn{5}{c}{$\log_{10}(\phi\,[\text{mag}^{-1}\,\text{Mpc}^{-3}])$} \\
        $M_\text{UV}$ & $z = 11$ & $z = 12$  & $z = 13$  & $z = 14$  & $z = 15$ \\
        \hline
        -22.0 & -7.39 & -8.36 & -10.03 & -10.21 & -12.18\\
        -21.5 & -6.56 & -7.51 & -8.82 & -10.15 & -12.04\\
        -21.0 & -5.99 & -6.79 & -7.90 & -9.81 & -10.82\\
        -20.5 & -5.56 & -6.31 & -7.30 & -9.51 & -10.33\\
        -20.0 & -5.14 & -5.83 & -6.73 & -9.07 & -10.08\\
        -19.5 & -4.78 & -5.41 & -6.16 & -8.61 & -9.16\\
        -19.0 & -4.44 & -5.13 & -5.76 & -7.75 & -8.85\\
        -18.5 & -4.17 & -4.70 & -5.31 & -6.92 & -8.50\\
        -18.0 & -3.85 & -4.37 & -4.97 & -6.30 & -7.91\\
        -17.5 & -3.60 & -4.11 & -4.57 & -5.66 & -7.53\\
        -17.0 & -3.29 & -3.80 & -4.25 & -5.32 & -7.05\\
        -16.5 & -3.14 & -3.52 & -3.96 & -4.80 & -6.63\\
        -16.0 & -2.85 & -3.29 & -3.74 & -4.43 & -6.36\\
        -15.5 & -2.66 & -3.03 & -3.46 & -4.21 & -5.98\\
        -15.0 & -2.43 & -2.84 & -3.20 & -3.81 & -5.64\\
        -14.5 & -2.27 & -2.61 & -3.03 & -3.57 & -5.20\\
        -14.0 & -2.12 & -2.40 & -2.70 & -3.36 & -4.61\\
        -13.5 & -1.90 & -2.28 & -2.60 & -3.08 & -4.19\\
        -13.0 & -1.75 & -2.04 & -2.28 & -2.84 & -3.92\\
        -12.5 & -1.55 & -1.90 & -2.14 & -2.67 & -3.49\\
        -12.0 & -1.41 & -1.65 & -1.94 & -2.35 & -3.18\\
        -11.5 & -1.23 & -1.51 & -1.75 & -2.32 & -2.92\\
        -11.0 & -1.02 & -1.28 & -1.60 & -1.98 & -2.70\\
        -10.5 & -0.89 & -1.11 & -1.43 & -1.78 & -2.55\\
        -10.0 & -0.76 & -0.97 & -1.17 & -1.57 & -2.20\\
        -9.5 & -0.57 & -0.84 & -1.00 & -1.36 & -1.98\\
        -9.0 & -0.55 & -0.71 & -0.91 & -1.25 & -1.88\\
        -8.5 & -0.76 & -0.95 & -1.08 & -1.33 & -1.68\\
        -8.0 & -0.83 & -1.05 & -1.16 & -1.35 & -1.67\\
        \hline
    \end{tabular}
\end{table}

\section{Comparison of Predicted and Observed UV Luminosity Functions}
\label{appendix:c}
\setcounter{table}{0} \renewcommand{\thetable}{C\arabic{table}}

\begin{figure}
    \includegraphics[width=\columnwidth]{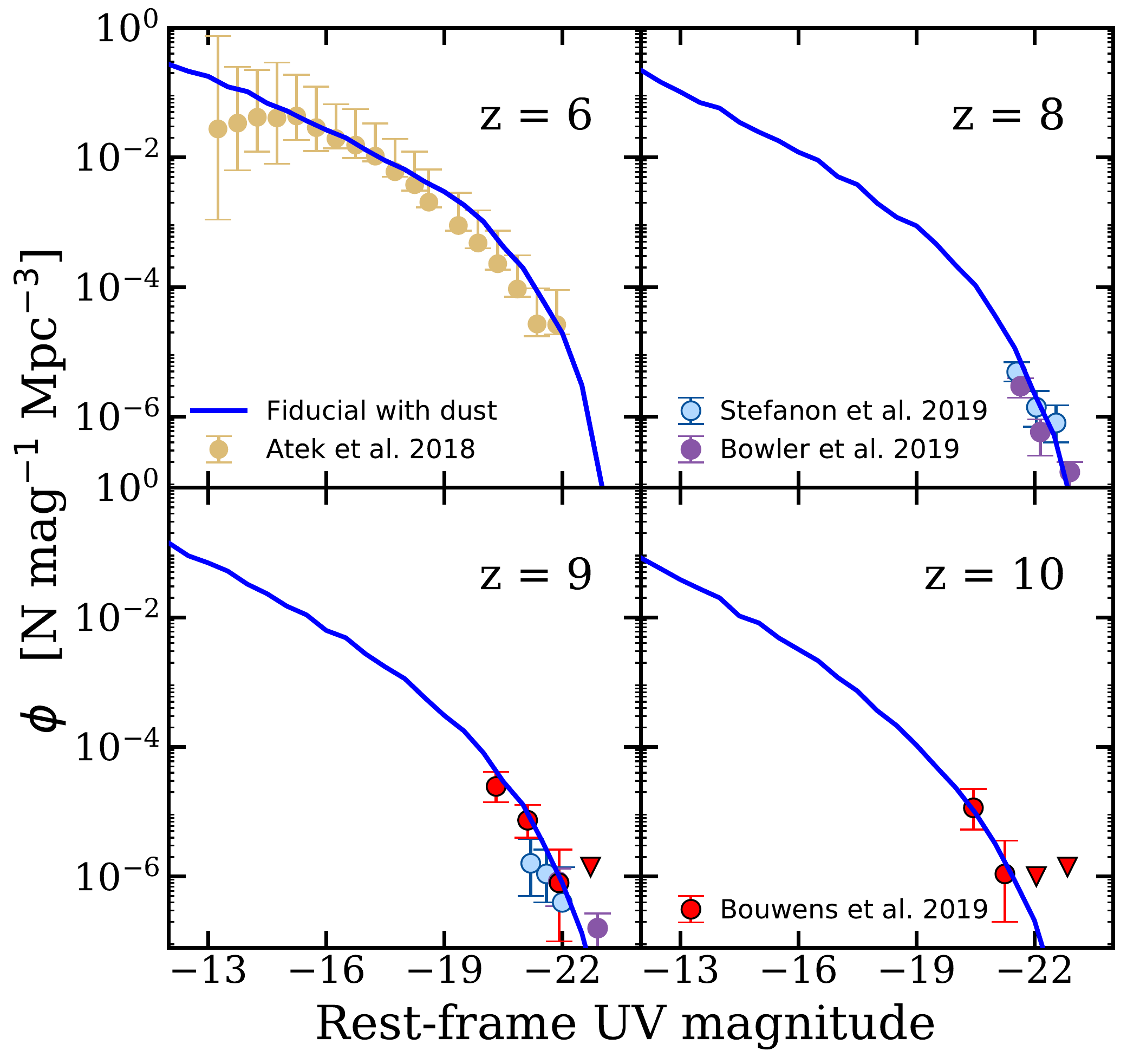}
    \caption{Rest-frame UV LFs from our fiducial model including dust attenuation, reproduced from \citetalias{Yung2019} and compared to the latest observational constraints from \citet[$z = 6$]{Atek2018}, \citet[$z = 8$ and 9]{Stefanon2019}, \citet[$z = 8$ and 9]{Bowler2020}, and \citet[$z = 9$ and 10]{Bouwens2019}. The agreement with these recent observations is excellent. }
    \label{fig:uvlf_newConstraints}
\end{figure}

In fig.~\ref{fig:uvlf_newConstraints}, we compare the predictions of our model to new observational constraints from \citet{Atek2018}, \citet{Stefanon2019}, and \citet{Bouwens2019}, which have been published in the interim since we first published the luminosity function predictions from our models. The models are exactly the same as those published in \citetalias{Yung2019}. The agreement with the new observations is excellent.

\bsp	
\label{lastpage}
\end{document}